\documentclass[12pt,letterpaper,pdftex]{article} 

\usepackage{algorithmic}
\usepackage{algorithm}
\usepackage{hyphenat}
\usepackage{lastpage}

\title{Transit-Length Distribution for Particle Transport in Binary Markovian Mixed Media}
\author{Brian C. Kiedrowski}

\usepackage{amsmath}
\usepackage{amsthm, amssymb}
\usepackage[margin=1in]{geometry}
\usepackage{setspace}
\usepackage{multirow}
\usepackage{textcomp}
\usepackage[pdftex]{graphicx}
\usepackage{sectsty}
\usepackage{subfig}
\usepackage{hyperref}
\usepackage{bm}

\usepackage{tikz}

\usetikzlibrary{decorations.pathmorphing}

\newcommand{\dir}{\hat{\mathbf{\Omega}}}

\newcommand{\dho}[2]{ \dfrac{ \partial {#1} }{ \partial {#2} } }

\newcommand{\adj}{\Psi^\dagger}
\newcommand{\iso}[2]{${}^{\scriptsize \textrm{#2}}${#1}}

\sectionfont{\normalsize}
\subsectionfont{\normalsize}
\subsubsectionfont{\normalsize}

\begin{document}


\doublespacing


%


%

\begin{center}
\noindent {\bf \large Transit-Length Distribution for Particle Transport in Binary Markovian Mixed Media}

\vspace{0.2in}

\noindent Brian C. Kiedrowski and Emily H. Vu \\
Department of Nuclear Engineering and Radiological Sciences, University of Michigan, 2355 Bonisteel Boulevard, Ann Arbor, MI 48109

\vspace{0.2in}


\vspace{0.5in}

\noindent
\textbf{Abstract} \\
\end{center}
The correspondence between the telegraph random process and transport within a binary stochastic Markovian mixture is established. This equivalence is used to derive the distribution function for the transit length, defined as the distance a particle moving along a straight-line trajectory travels through a specific material zone within the random mixture. A numerically robust asymptotic form of this distribution is obtained for highly mixed materials and the convergence to the atomic-mix limit is shown. The validity of the distribution is verified using a Monte Carlo simulation of the transport process. The distribution is applied to particle transport in slab geometry containing porous media for two cases: the transmission of light and the stopping of charged particles. For both of these applications, analytical forms using the approximate asymptotic model for the transmission probability of beam sources are obtained and illustrative numerical results are provided. These results show that in cases of highly mixed materials, the asymptotic forms are more accurate than the atomic-mix limit.
\section{Introduction} \label{Sec:Introduction}

Accurately describing the transport of particle radiation through porous or turbid media requires modeling its underlying stochasticity. The naive description is the atomic-mix model, which is a simple volumetric homogenization, but this crude model often proves too inaccurate for applications. As such, several researchers have devised more faithful mathematical descriptions incorporating the material randomness. Levermore et al~\cite{Levermore_TransportRandomMedia_JMathPhys27_1986} derived the Levermore-Pomraning closure for the ensemble-averaged distribution of radiation under the assumption of a binary mixture with exponentially distributed, i.e. Markovian, chord lengths that provides a more faithful description of random media than atomic mix. Since then, several researchers including Levermore and colleagues~\cite{Levermore_RenewalTheoryTransport_JMathPhys29_1988}, Sanchez~\cite{Sanchez_LinearKineticTheoryStochasticMedia_JMathPhys30_1989}, Davis~\cite{Davis_GeneralizedLinearTransportCorrelatedStochastic_JCTT43_2014}, Pautz and Franke~\cite{Pautz_GeneralizedLPClosure_MC2015_2015}, and d'Eon~\cite{dEon_RecriprocaFormulationBinaryMarkovian_arXiv_2019} have developed more general mathematical descriptions. 



In this article we model binary Markovian stochastic media using a system of first-order partial differential transport equations relating two state variables, the total distance traveled within a material and the difference of the distances between the two different materials, that is identical to the telegrapher's equations. We use the solutions of this system of differential equations to obtain the transit-length distribution, the probability density of the distance that a particle travels along a straight trajectory through a particular material type given that the starting material type is known.

The telegrapher's equations represent a random process describing a particle traveling at two fixed velocities and switching directions at exponentially distributed times. The connection of the telegrapher's equations to this random process was first formalized by Kac~\cite{Kac_StochasticModelTelegrapher_RMJMath4_1974}. A result similar to the transit-length density obtained in this article was derived by Ratanov~\cite{Ratanov_JumpTelegraphModel_QuantFinance7_2007} in the context of quantitative finance for the occupation time of a particular state of a telegraph process. To our knowledge, however, this is the first time the telegraph process has been applied to transport in stochastic media, with the novel insight of this research being the observation of an equivalence between the temporal and spatial variables of a standard telegraph process to the previously mentioned state variables and the switching of directions to the transition between material types. In our related conference summaries~\cite{Vu_AngularDeflectionCondensedHistory_ANS125_2021,Vu_EnergyLossCondensedHistory_ANS125_2021}, we have given this resulting transit-length distribution but without the derivation that we provide in this article and applied it differently as a weighting function to homogenize material properties of binary Markovian mixtures to produce distribution functions that efficiently model the transport of electrons.

The transit-length distribution involves modified Bessel functions of the first kind that evaluate to very large numbers for media with a high degree of mixing, leading to numerical overflow. To address this, we derive an analytical asymptotic form of the transit-length distribution that is numerically stable in this highly mixed case. Using this asymptotic form, we then show that the transit-length distribution satisfies the atomic-mix limit. We provide implementations in C++, Python, and Matlab in a public software repository\footnote{\url{https://github.com/bckiedrowski/transit_dist}} that also includes all the scripts used to generate the results within this article. Then, we show results from a Monte Carlo direct simulation of the transport process that provide evidence that the mathematical model is consistent and has been correctly implemented. We also give numerical evidence that the asymptotic and exact forms converge in the limit of large mixing.

We then apply the transit-length distribution to the special case of porous media, which consists of a binary mixture of some material and void, in slab geometry. First, we consider the transport of light, but the results are applicable to any neutral particle where the material is effectively a pure absorber. We obtain a form of the expected, i.e. ensemble-average of the angular flux in terms of the transit-length distribution. This angular flux must be integrated numerically using the exact form of the transit-length distribution. However, we obtain an approximate analytical form for the angular flux using the asymptotic form that is valid when the material is not too optically thick. We also integrate these expressions numerically to determine the transmission probability, or equivalently the outgoing partial current. We provide numerical results that illustrate the behavior of these quantities and quantify the accuracy of the asymptotic form and the atomic-mix limit. The results show that when the conditions for its validity are met (highly mixed and not too optically thick), the analytical result using the approximate asymptotic form is more accurate than the result obtained with the atomic-mix model.

Next, we show results for the slowing down or stopping of charged particles. (In this paper we only assume the continuous slowing down approximation and neglect stochastic effects of energy straggling and angular deflection, which we treat in other conference publications \cite{Vu_AngularDeflectionCondensedHistory_ANS125_2021,Vu_EnergyLossCondensedHistory_ANS125_2021}.) We derive expressions for the transmission probability and energy spectrum. For the former, we also apply the asymptotic form of the transit-length distribution and discuss the limitations of its applicability in this problem. We then provide illustrative numerical results for electrons incident onto a porous slab of iron.

\section{Transit-Length Distribution} \label{Sec:TransitLengthDistribution}
This section gives the derivation of the probability distribution for the transit length, the amount of distance a particle moving through a binary Markovian stochastic mixture along a straight trajectory is within a particular material type, given that it started within a prescribed material type. Asymptotic forms are obtained for the case where the number of zones traversed is large and the atomic-mix limit of this distribution is demonstrated. Considerations for a practical implementation are discussed. Finally, the distribution and implementation is verified using Monte Carlo sampling.

\subsection{Derivation} \label{Sec:TransitLengthDistribution_Derivation}
The problem being considered is a random mixture of two different materials, $A$ and $B$. The distance a particle would travel along a straight trajectory within a region containing a particular material or stochastic zone is randomly distributed and given by an exponential distribution characterized by a mean chord length:
\begin{align} 
  \Lambda_A \equiv 	&\text{ The mean distance a particle moving along straight line within material} \nonumber \\
             		&\text{ type $A$ travels to cross into an adjacent zone of material type $B$.}   \nonumber
\end{align}
The mean chord length $\Lambda_B$ has the equivalent definition with $A$ and $B$ switched. To simplify the notation going forward, we define
\begin{subequations}
\begin{align}
  \alpha &= \frac{1}{\Lambda_A}, \\
  \beta  &= \frac{1}{\Lambda_B},
\end{align}
\end{subequations}
which are the rate coefficients per unit distance of a particle transitioning between the respective regions.

Next, let $s$ be the total distance that a particle travels across the mixture in either region. We now define the following random variable:
\begin{align}
  X \equiv 	&\text{ The distance traveled in material type $A$ minus the distance traveled in material} \nonumber \\
  			&\text{ type $B$ given that a total distance $s$ has been traveled.} \nonumber 
\end{align}
This can be thought of as the random net distance spent within material type $A$. A positive value denotes that more distance has been traversed in material $A$, whereas a negative value means that more distance has been traversed in material $B$. We will often refer to this in short as the net distance. We further condition this on the material type the particle begins in as $X_A$ and $X_B$.

Next, we define the random variables for the transit lengths:
\begin{align}
  L_{A|A} \equiv	&\text{ The total distance traveled in material type $A$ given that the particle} \nonumber \\
  					&\text{ began in material type $A$,} \nonumber \\
  L_{A|B} \equiv	&\text{ The total distance traveled in material type $A$ given that the particle} \nonumber \\
  					&\text{ began in material type $B$.} \nonumber
\end{align}
The random variables for $L_{B|A}$ and $L_{B|B}$ can be defined similarly, and are related parametrically to the total traveled distance $s$ by
\begin{subequations}
\begin{align}
  L_{B|A} = s - L_{A|A}, \\
  L_{B|B} = s - L_{A|B}. 
\end{align}
\end{subequations}
We can relate this to the net distance conditional upon the starting material by
\begin{subequations}
\begin{align}
  X_A = L_{A|A} - L_{B|A} = 2 L_{A|A} - s, \label{Eq:relating_XA_LAA} \\
  X_B = L_{A|B} - L_{B|B} = 2 L_{A|B} - s. \label{Eq:relating_XB_LAB}
\end{align}
\end{subequations}

Given these definitions, we develop transport problems for a probability density $p_A^m(x,s)$, the probability per unit net distance $x$ given total distance $s$ for a particle traveling in material $A$ having crossed $m$ random material zones. The solution of these transport problems can then be added together as a Neumann expansion to obtain probability densities for the transit lengths. 

This sum or superposition of all possible transitions is mathematically equivalent to the telegraph process, which describes a particle that travels with some speed along the $x$ axis in either the positive or negative direction and switches at random intervals. In this context, the total distance $s$ has the role of a time-like variable, the net distance $x$ has the role of position, the speed is unity, and $\alpha$ and $\beta$ are the rate coefficients for transitions at exponentially distributed times. In the case of a typical telegraph process, the particle switches direction; whereas in this application, these switches correspond to transitions from one material zone to the other.

To begin the mathematical model, we start by writing partial differential equations for the base case for particles that have had zero crossings starting in material types $A$ and $B$ respectively:
\begin{subequations}
\begin{align}
  &\dho{p_A^0}{s} + \dho{p_A^0}{x} + \alpha p_A^0(x,s) = 0, \quad p_A^0(x,0) = \delta(x); \\
  &\dho{p_B^0}{s} - \dho{p_B^0}{x} + \beta  p_B^0(x,s) = 0, \quad p_B^0(x,0) = \delta(x).
\end{align}
\end{subequations}
The domain of this and all subsequent transport problems is $-\infty < x < \infty$, $s \ge 0$. The initial condition or source term is a Dirac delta function denoting the particle begins with a unit source at $x = 0$ with $s = 0$, no distance traveled initially. The $x$ partial derivative terms ensure that as $s$ increases, $x$ increases or decreases, respective to the starting region, at the same rate. The third term is an absorption-like term for particles undergoing their first transition.

These equations are uniform transport problems with absorption and can be solved using the method of characteristics. This yields:
\begin{subequations}
\begin{align}
  &p_A^0(x,s) = \delta( x - s ) e^{-\alpha s}, \\
  &p_B^0(x,s) = \delta( x + s ) e^{-\beta s} .
\end{align}
\end{subequations}

Given the base case, we can now write down partial differential equations for $m > 0$ transitions for particles within material types $A$ and $B$ respectively:
\begin{subequations}
\begin{align}
  &\dho{p_A^m}{s} + \dho{p_A^m}{x} + \alpha p_A^m(x,s) = \beta  p_B^{m-1}(x,s), \quad p_A^m(x,0) = 0; \\
  &\dho{p_B^m}{s} - \dho{p_B^m}{x} + \beta  p_B^m(x,s) = \alpha p_A^{m-1}(x,s), \quad p_B^m(x,0) = 0.
\end{align}
\end{subequations}
These equations for particles having at least one transition differ in that there is not an inhomogeneous source term from particles that have just underwent their $(m - 1)$th transition with a zero initial condition since particles normally do not begin with any transitions. To solve these equations, we introduce the transformations
\begin{subequations}
\begin{align}
  &p_A^m(x,s) = \tilde{p}_A^m(x,s) \exp \left[ -\alpha s + \nu ( s - x ) \right] , \label{Eq:MOCtransform_pA} \\
  &p_B^m(x,s) = \tilde{p}_B^m(x,s) \exp \left[ -\beta  s - \nu ( s + x ) \right] . \label{Eq:MOCtransform_pB}
\end{align}
with
\begin{align}
  \nu = \frac{ \alpha - \beta }{2}. 
\end{align}
\end{subequations}
This allows the equations to be rewritten in terms of $\tilde{p}_A^m(x,s)$ and $\tilde{p}_B^m(x,s)$, eliminating the absorption terms:
\begin{subequations}
\begin{align}
  &\dho{\tilde{p}_A^m}{s} + \dho{\tilde{p}_A^m}{x} = \beta  \tilde{p}_B^{m-1}(x,s), \quad \tilde{p}_A^m(x,0) = 0; \label{Eq:transform_pAm} \\
  &\dho{\tilde{p}_B^m}{s} - \dho{\tilde{p}_B^m}{x} = \alpha \tilde{p}_A^{m-1}(x,s), \quad \tilde{p}_B^m(x,0) = 0. \label{Eq:transform_pBm} 
\end{align}
Additionally,
\begin{align}
  \tilde{p}_A^0(x,s) &= \delta( x - s ) e^{-\nu ( s - x ) }, \label{Eq:transform_pA0} \\
  \tilde{p}_B^0(x,s) &= \delta( x + s ) e^{-\nu ( s + x ) }. \label{Eq:transform_pB0} 
\end{align}
\end{subequations}

Now we construct the probability density for $X_A$, the net distance given the particle started in material type $A$, by recursively solving Eqs.~\eqref{Eq:transform_pAm} and~\eqref{Eq:transform_pBm} for successive number of transitions and then summing the result. We begin with Eq.~\eqref{Eq:transform_pA0} for zero transitions and note that solution exists on the line $s = x$ in the $x$-$s$ plane. 

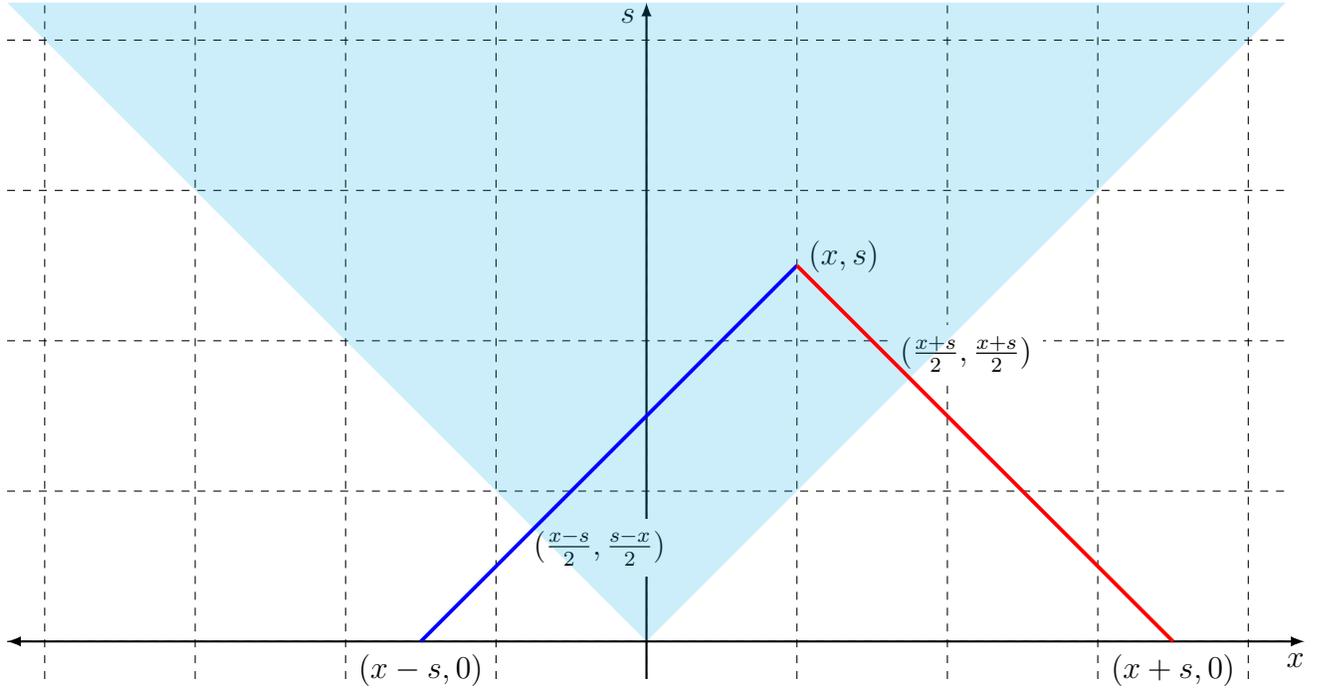
\begin{figure}[t!]
\begin{center}
\begin{tikzpicture}
  \draw[-latex,thick] (0,0) -- (8.75,0);
  \draw[-latex,thick] (0,0) -- (-8.5,0);
  \node at (8.625,-0.25) {$x$};
  \draw[-latex,thick] (0,-0.5) -- (0,8.5);
  \node at (-0.25,8.325) {$s$};
  \foreach \x in {-8,-6,-4,-2,0,2,4,6,8}
    \draw[dashed] (\x,-0.5) -- (\x,8.5);
  \foreach \y in {0,2,4,6,8}
    \draw[dashed] (-8.5,\y) -- (8.5,\y);
  \node[fill=white] at (2.625,5.125) {$(x,s)$};
  \node at (-3,-0.375) {$(x-s,0)$};
  \node at ( 7,-0.375) {$(x+s,0)$};
  \node[fill=white] at (-0.625,1.25) {$(\frac{x-s}{2},\frac{s-x}{2})$};
  \node[fill=white] at ( 4.25,3.825) {$(\frac{x+s}{2},\frac{x+s}{2})$};

  \fill[fill=cyan, opacity=0.2] (0,0) -- (8.5,8.5) -- (-8.5,8.5) -- cycle;
  \draw[blue,line width=0.5mm] (-3,0) -- (2,5);
  \draw[red, line width=0.5mm] ( 7,0) -- (2,5);
\end{tikzpicture}
\caption{Illustration of characteristics on the $x$-$s$ plane.}
\label{Fig:characteristics}
\end{center}
\end{figure}

The probability densities for subsequent material transitions are solved using the method of characteristics. Figure~\ref{Fig:characteristics} illustrates the domain and chracteristic lines for a particle with coordinate $(x,s)$. The shaded area denotes the meaningful range of $(x,s)$, where $s > x$ and $s > -x$. (Points outside this domain are not attainable because that would require the magnitude of the net distance to exceed the total distance.) The other coordinates in Fig.~\ref{Fig:characteristics} are with respect to $(x,s)$ and denote the points the characteristics cross the $x$ axis (the formal start of the characteristic) and where they cross the meaningful shaded domain. For the case where the particle begins in material type $A$, an odd number of transitions uses characteristics that are diagonal lines (at $45^\circ$ and given in red in Fig.~\ref{Fig:characteristics}) pointing up and to the left with $x$ decreasing as $s$ increases; an even number of transitions uses characteristics that are diagonal lines but pointing up and to the right (blue line in Fig.~\ref{Fig:characteristics}) with $x$ increasing as $s$ increases. 

Next, we solve for transformed probability density for a single transition from material $A$ into $B$ along a characteristic to obtain
\begin{align}
  \tilde{p}_B^1(x,s) = \alpha \int_0^s \tilde{p}_A^0(x + s - t,t) dt. \nonumber
\end{align}
Here the functional dependencies are such that the characteristic is parameterized with $s(t) = t$ and the red characteristic line being given by $x(t) = x + s - t$. Since $\tilde{p}_A^0$ in Eq.~\eqref{Eq:transform_pA0} contains a Dirac delta function at the line $s = x$ on the edge of the meaningful domain, the integral will only be nonzero for values of $s$ in this domain. Carrying this out yields
\begin{align}
  \tilde{p}_B^1(x,s) = \frac{\alpha}{2} \Theta( (s - x)(s + x) ) , \label{Eq:transform_pB1} 
\end{align}
where $\Theta(x)$ is the Heaviside step function that is one when the argument is non-negative and zero otherwise.

Now we can solve for the probability density following the second transition as
\begin{align}
  \tilde{p}_A^2(x,s) = \beta \int_0^s \tilde{p}_B^1(x - s + t,t) dt. \nonumber
\end{align}
This uses the blue characteristic line given by $s(t) = t$ and $x(t) = x - s + t$. The presence of the Heaviside step function ensures that all points in the integrand for $t < (s-x)/2$ are zero. Therefore, we can rewrite the integrand without the step function by changing the integration bounds and then removing it from the integral since it is then a constant term. Evaluating the integral gives
\begin{align}
  \tilde{p}_A^2(x,s) 
  &= \frac{\alpha \beta}{2} \Theta( (s - x)(s + x) ) \int_{\frac{s-x}{2}}^s  dt \nonumber \\ 
  &= \frac{\alpha \beta}{4} ( s + x ) \Theta( (s - x)(s + x) ) . \label{Eq:transform_pA2} 
\end{align}

Continuing with the third transition, we have
\begin{align}
  \tilde{p}_B^3(x,s) 
  &= \alpha \int_0^s \tilde{p}_A^2(x + s - t,t) dt \nonumber \\
  &= \frac{ \alpha^2 \beta }{4} \Theta( (s - x)(s + x) ) \int_{\frac{s+x}{2}}^s ( s + x ) dt \nonumber \\
  &= \frac{ \alpha^2 \beta }{8}  ( s + x ) ( s - x ) \Theta( (s - x)(s + x) ) .
\end{align}
This again follows the red characteristic line for odd transitions.

The last couple integrals illustrate the idea for higher-order transitions, from which a pattern can be observed. These involve two kinds of integrals. One for odd numbers of transitions and another for even numbers. These are, respectively,
\begin{subequations}
\begin{align}
  \int_{\frac{s+x}{2}}^s ( 2t - x - s )^{n-1} ( s + x )^n dt &= \frac{1}{2} \frac{ ( s + x )^n ( s - x )^n }{ n } , \\
  \int_{\frac{s-x}{2}}^s ( 2t + x - s )^n     ( s - x )^n dt &= \frac{1}{2} \frac{ ( s + x )^{n+1} ( s - x )^n }{ n + 1 } .  
\end{align}
\end{subequations}
Here $n$ is the $n$th \emph{positive} odd or even transition; zero is excluded because it has a fundamentally different form, i.e. involving a Dirac delta function.

Given this indexing and these integrals, we can write the transformed probability density for the net distance $x$ given that the particle begins in material type $A$. This is done by taking the sum of the terms for zero transitions, those having had an odd number of transitions, and those with an even number of transitions (excluding zero):
\begin{align}
  \tilde{p}_{X_A}(x,s) &= \tilde{p}_A^0(x,s) + \sum_{n=0}^\infty \tilde{p}_B^{2n+1}(x,s) +  \sum_{n=0}^\infty \tilde{p}_A^{2n+2}(x,s) . 
\end{align}
Inserting the results we have,
\begin{align}
  \tilde{p}_{X_A}(x,s) &= \delta( x - s ) e^{-\nu ( s - x ) } + \Theta( (s - x)(s + x) ) \nonumber \\  &\times \left[ \frac{\alpha}{2} \sum_{n=0}^\infty \frac{ \alpha^n \beta^n }{ 2^{2n} (n!)^2} ( s + x )^n ( s - x )^n 
  + \frac{1}{2} \sum_{n=0}^\infty \frac{ \alpha^{n+1} \beta^{n+1} }{ 2^{2n+1} n! (n+1)! } ( s + x )^{n+1} ( s - x )^n \right].
\end{align}
Through a bit of manipulation, the two summations can be written in terms of modified Bessel functions of the first kind, which are defined via the following power series:
\begin{subequations}
\begin{align}
  I_0(z) &= \sum_{n=0}^\infty \frac{z^{2n}}{2^{2n} (n!)^2}, \\
  I_1(z) &= \sum_{n=0}^\infty \frac{z^{2n+1}}{2^{2n+1} n!(n+1)!} 
\end{align}
\end{subequations}
Recasting in terms of these modified Bessel functions gives
\begin{align}
  \tilde{p}_{X_A}(x,s) &= \delta( x - s ) e^{-\nu ( s - x ) } + \Theta( (s - x)(s + x) ) \nonumber \\ &\times \left[ \frac{\alpha}{2} I_0 \left( \sqrt{\alpha \beta (s + x)(s - x) } \right) 
  + \frac{1}{2} \sqrt{ \frac{ \alpha \beta ( s + x ) }{ s - x } } I_1 \left( \sqrt{\alpha \beta (s + x)(s - x) } \right) \right].
\end{align}
Finally, this transformed probability density function is put in terms of the standard version using Eq.~\eqref{Eq:MOCtransform_pA}. This gives the result
\begin{align}
  p_{X_A}(x,s) &= \delta( x - s ) e^{-\alpha s } + \Theta( (s - x)(s + x) )  e^{-\alpha s + \nu ( s - x ) }\nonumber \\ 
  &\times \left[ \frac{\alpha}{2} I_0 \left( \sqrt{\alpha \beta (s + x)(s - x) } \right) 
  + \frac{1}{2} \sqrt{ \frac{ \alpha \beta ( s + x ) }{ s - x } } I_1 \left( \sqrt{\alpha \beta (s + x)(s - x) } \right) \right]. \label{Eq:pXA}
\end{align}

Following a very similar analysis, the probability density function for the net distance given that the particle begins in material type $B$ is obtained. For this case,
\begin{align}
  \tilde{p}_{X_B}(x,s) &= \tilde{p}_B^0(x,s) + \sum_{n=0}^\infty \tilde{p}_A^{2n+1}(x,s) +  \sum_{n=0}^\infty \tilde{p}_B^{2n+2}(x,s) . 
\end{align}
These are the $\tilde{p}_A^m$ and $\tilde{p}_B^m$ that were not used to calculate $p_{X_A}$. To evaluate them, we first use the base case from Eq.~\eqref{Eq:transform_pB0}. Then the others can be evaluated using the following two integrals,
\begin{subequations}
\begin{align}
  \int_{\frac{s-x}{2}}^s ( 2t + x - s )^{n-1} ( s - x )^n dt &= \frac{1}{2} \frac{ ( s - x )^n ( s + x )^n  }{ n } , \\
  \int_{\frac{s+x}{2}}^s ( 2t - x - s )^n     ( s + x )^n dt &= \frac{1}{2} \frac{ ( s - x )^{n+1} ( s + x )^n }{ n + 1 } ,
\end{align}
\end{subequations}
for odd numbers of transitions and even numbers of transitions, respectively. After evaluating the integrals, these can be cast in terms of the modified Bessel functions and transformed back to obtain the final result:
\begin{align}
  p_{X_B}(x,s) &= \delta( x + s ) e^{-\beta s } + \Theta( (s - x)(s + x) )  e^{-\beta s - \nu ( s + x ) }\nonumber \\ 
  &\times \left[ \frac{\beta}{2} I_0 \left( \sqrt{\alpha \beta (s + x)(s - x) } \right) 
  + \frac{1}{2} \sqrt{ \frac{ \alpha \beta ( s - x ) }{ s + x } } I_1 \left( \sqrt{\alpha \beta (s + x)(s - x) } \right) \right]. \label{Eq:pXB}
\end{align}

Now that we have the probability densities for the net distance given that the particle begins in materials $A$ or $B$, we proceed with finding the transit-length distributions using a simple transformation of variables using Eqs.~\eqref{Eq:relating_XA_LAA} and~\eqref{Eq:relating_XB_LAB}:
\begin{subequations}
\begin{align}
  f_{A|A}(\ell,s) = 2 p_{X_A} ( 2 \ell - s , s ), \\
  f_{A|B}(\ell,s) = 2 p_{X_B} ( 2 \ell - s , s ),
\end{align}
\end{subequations}
where $\ell$ is the transit length in material type $A$. Carrying through this transformation, we arrive at the probability densities for the total distance traveled in material $A$ given that the particle begins in material types $A$ and $B$ respectively,
\begin{subequations}
\begin{align}
  f_{A|A}(\ell,s) &= \delta( \ell - s ) e^{-\alpha s} + \Theta( \ell ( \ell - s ) ) e^{-\alpha \ell - \beta ( s - \ell )} \nonumber \\
  &\times \left[ \alpha I_0 \left( 2 \sqrt{\alpha \beta \ell (s - \ell ) } \right) + \sqrt{ \frac{ \alpha \beta  \ell }{ s - \ell } } I_1 \left( 2 \sqrt{\alpha \beta \ell (s - \ell ) } \right) \right] , \label{Eq:TransitLengthDensity_AA} \\
  f_{A|B}(\ell,s) &= \delta( \ell) e^{-\beta s} + \Theta( \ell ( \ell - s ) ) e^{-\alpha \ell - \beta ( s - \ell )} \nonumber \\
  &\times \left[ \beta I_0 \left( 2 \sqrt{\alpha \beta \ell (s - \ell ) } \right) + \sqrt{ \frac{ \alpha \beta ( s - \ell ) }{ \ell } } I_1 \left(  2 \sqrt{\alpha \beta \ell (s - \ell ) } \right) \right] . \label{Eq:TransitLengthDensity_AB}
\end{align}
\end{subequations}

The distribution for the distance in material type $A$ where the probabilities of starting types $A$ and $B$ are given by $p_A = \omega$ and $p_B = 1 - \omega$ respectively is the linear combination of the densities:
\begin{align}
  f_A(\ell,s) = \omega f_{A|A}(\ell,s) + ( 1 - \omega ) f_{A|B}(\ell,s) . 
\end{align}
Expanding this out gives
\begin{align}
  f_A(\ell,s) &= \omega \delta( \ell - s ) e^{-\alpha s} + ( 1 - \omega ) \delta( \ell ) e^{-\beta s } + \Theta( \ell ( \ell - s ) ) e^{-\alpha \ell - \beta ( s - \ell )} \nonumber \\
   &\times \Bigg[ \bigg( \omega \alpha + ( 1 - \omega ) \beta \bigg) I_0 \left( 2 \sqrt{\alpha \beta \ell (s - \ell ) } \right) \nonumber \\
  &\quad + \sqrt{ \alpha \beta } \left( \frac{ ( 2 \omega - 1 ) \ell + ( 1 - \omega ) s }{ \sqrt{ \ell ( s - \ell ) } } \right) I_1 \left( 2 \sqrt{\alpha \beta \ell (s - \ell ) } \right) \Bigg] . \label{Eq:TransitLengthDensity_A}
\end{align}
One numerical issue is that the boundary points, $\ell = 0$ and $\ell = s$, are in an indeterminate form of $0/0$ because of the $I_1$ term. Special forms for numerical evaluation of the density part (excluding the Dirac delta terms) at these points can be derived using L'Hopital's rule. These are
\begin{subequations}
\begin{align}
  \lim_{\ell \rightarrow 0^+} f_A(\ell,s) &= \left[ \omega \alpha + ( 1 - \omega ) \beta ( 1 + \alpha s ) \right] e^{-\beta s}  , \label{Eq:TransitLengthDensity_A_limit_0} \\
  \lim_{\ell \rightarrow s^-} f_A(\ell,s) &= \left[ ( 1 - \omega ) \beta + \omega \alpha ( 1 +  \beta s ) \right] e^{-\alpha s} . \label{Eq:TransitLengthDensity_A_limit_s}
\end{align}
\end{subequations}

Analogous probability density functions for the distance traveled in material type $B$ can be obtained similarly with a variable transformation since it is parametrically given by $s - \ell$. As before, in these equations, the Dirac delta function term corresponds to particles that have not transitioned from their starting material, the $I_0$ term corresponds to particles that have undergone an odd number of transitions, and the $I_1$ term corresponds to those that have had a positive, even number of transitions.

\subsection{Asymptotic Forms} \label{Sec:TransitLengthDistribution_AsymptoticForms}

The transit-length distributions in Eqs.~\eqref{Eq:TransitLengthDensity_AA} and~\eqref{Eq:TransitLengthDensity_AB} are exact results. However, as the expected number of regions traversed within distance $s$ becomes large, the arguments of the modified Bessel functions of the first kind become large, causing numerical overflow. This motivates the derivation of asymptotic forms that are more amenable to numerical calculations when the arguments of the modified Bessel functions become large. From these forms, we demonstrate that the transit-length distribution satisfies the theoretical atomic mix limit in the following section.

To begin, we parameterize $\beta$ in terms of $\alpha$ as
\begin{align}
  \beta = k \alpha, \quad k \gtrsim 1.
\end{align}
We require that $\alpha$ is large, and since $k$ is on the order of or greater than 1, $\beta$ is large as well. Next, the modified Bessel function of any order has the following form for large argument:
\begin{align}
  I_n(z) \sim \frac{ e^z }{ \sqrt{ 2 \pi z } }, \quad z \rightarrow \infty . \label{Eq:Bessel_Asymptotic}
\end{align}
In this context, we let
\begin{align}
  z =  2 \sqrt{\alpha \beta \ell (s - \ell ) } = 2 \alpha \sqrt{k \ell (s - \ell ) } . \label{Eq:Bessel_zDefinition}
\end{align}
Since $\alpha$ is large, $z$ is also large. Applying the asymptotic form for the modified Bessel functions of the first kind and the above definitions on Eq.~\eqref{Eq:TransitLengthDensity_AA} and grouping the exponentials gives
\begin{align}
  \tilde{f}_{A|A}(\ell,s) =  C_A \left[ 1 + \sqrt{\frac{k \ell }{s - \ell}} \right] 
  \frac{\alpha}{2 \sqrt{\pi \alpha \sqrt{ k \ell ( s - \ell ) } }} \exp \left[ -\alpha \ell - k \alpha ( s - \ell ) 2 \alpha \sqrt{ k \ell ( s - \ell ) } \right]  .
\end{align}
Here the range is taken from $0 \le \ell \le s$. The Dirac delta term vanishes in the limit of large $\alpha$. 

An important point is that the application of asymptotic form of the modified Bessel function in Eq.~\eqref{Eq:Bessel_Asymptotic} does not preserve the area under the curve, or, in this context, total probability. Therefore, a renormalization factor $C_A$ is introduced. Note that this factor approaches one in the limit that $\alpha \rightarrow \infty$. Additionally, the action of replacing the Bessel functions with the asymptotic form leads to an equation where the densities are infinite at the boundary values of 0 and $s$ on the count of the $( \ell ( s - \ell ) )^{-1/4}$ term; however, the integral under the curve is still finite and therefore the distribution is normalizable. Furthermore, the arguments of the Bessel functions approach zero at the endpoints, and are therefore small, so using Eq.~\eqref{Eq:Bessel_Asymptotic} in that portion of the domain is questionable; however, the contribution of the boundaries to the integral is negligible under the stated conditions, so even if this model is used over the entire domain, any error is very minor, and the results shown later in this paper demonstrate this.

The first term in brackets can be given a common denominator and the exponential term can be factored to obtain
\begin{align}
  \tilde{f}_{A|A}(\ell,s) =  \frac{C_A}{2} \left[ \frac{ \sqrt{s-\ell} + \sqrt{k\ell} }{ \sqrt{s-\ell} } \right] 
  \sqrt{ \frac{\alpha}{\pi \sqrt{ k \ell ( s - \ell ) } } } \exp \left[ -\alpha \left( \sqrt{\ell} - \sqrt{ k (s - \ell) } \right)^2 \right]  .
\end{align}
Next, we introduce
\begin{subequations}
\begin{align}
  u &= \sqrt{\ell} - \sqrt{k(s-\ell)}, \label{Eq:asymptoticForm_uTransformation_Definition} \\
  \ell(u) &= \left[ \frac{ u + \sqrt{ k ( 1 + k) s - k u^2 } }{ 1 + k } \right]^2 ,
\end{align}
and transform $\ell \rightarrow u$ with
\begin{align}
  \frac{d\ell}{du} = \frac{ 2 \sqrt{ \ell ( s - \ell ) } }{ \sqrt{ s - \ell } + \sqrt{ k \ell } } 
\end{align}
\end{subequations}
to obtain
\begin{align}
  \tilde{f}_{A|A}(u,s) = C_A \left[ \frac{\ell(u) }{ k ( s - \ell(u) ) } \right]^{1/4}  \sqrt{ \frac{\alpha}{\pi} } e^{-\alpha u^2}, \quad -\sqrt{ks} \le u \le \sqrt{s}. \label{Eq:fAA_asymptoticSqrt}
\end{align}
Observe that the term on the right is a Gaussian in $u$ with mean zero and variance $1/(2\alpha)$. The term on the left adds a positive skew that diminishes as $\alpha$ gets large. This skew is the result of the particle starting within material type $A$. Also of note is the transit length $\ell$ at $u = 0$:
\begin{align}
  \ell(0) = \frac{ k s }{ 1 + k } = \frac{ \Lambda_A }{ \Lambda_A + \Lambda_B } s. \label{Eq:l_at_u_equal_zero}
\end{align}
This is the expected transit length in the atomic-mix limit.

Equation~\eqref{Eq:fAA_asymptoticSqrt} is an asymptotic form that is numerically stable over a much larger range of values than the exact form. The normalization constant $C_A$, however, cannot be determined analytically. We can, however, introduce one further approximation to yield a form that is analytic in terms of standard functions. We take first-order Taylor series expansion of the one-fourth root term of in $u$ about $u = 0$:
\begin{align}
  f^*_{A|A}(u,s) = C_A \left[ 1 + \frac{1}{2} \sqrt{ \frac{1+k}{ks} } u \right] \sqrt{ \frac{\alpha}{\pi} } e^{-\alpha u^2}, \quad -\sqrt{ks} \le u \le \sqrt{s}. \label{Eq:fAA_asymptoticLinear}
\end{align}
The loss of accuracy from doing this expansion is minimal so long as the peak of the Gaussian is narrow enough so that the value of the distribution is very small near the extrema. If this is not the case, it violates the initial assumption of the particle crossing numerous regions. The constant $C_A$ is again taken to be an appropriate renormalization coefficient for Eq.~\eqref{Eq:fAA_asymptoticLinear} such that integrating over $u$ from $-\sqrt{ks}$ to $\sqrt{s}$ gives unity. The normalization constant for Eq.~\eqref{Eq:fAA_asymptoticLinear} is then
\begin{align}
C_A = 2 \left[ \operatorname{erf}\left(\sqrt{k\alpha s}\right)+\operatorname{erf}\left(\sqrt{\alpha s}\right) + \frac{1}{2} \sqrt{ \frac{1+k}{ks} } \cdot \dfrac{ e^{-k \alpha s}  - e^{-\alpha s}  }{\sqrt{{\pi \alpha}}} \right]^{-1} . \label{Eq:fAA_asymptoticLinear_NormalizationCoefficient}
\end{align}

Equation~\eqref{Eq:fAA_asymptoticLinear} yields a negative probability density within the range when $k > 3$. This point occurs for $u < -2 \sqrt{ks/(k+1)}$. In practice, this is outside the domain of applicability of the asymptotic model. Even so, the Gaussian is very small when this occurs for cases where one would apply the asymptotic model and the negative contribution to the area for finding $C_A$ is negligible. One could truncate the range in Eq.~\eqref{Eq:fAA_asymptoticLinear} to only include the positive area and then rederive Eq.~\eqref{Eq:fAA_asymptoticLinear_NormalizationCoefficient} with the new lower limit of integration. We do not do this in this particular case, but rather handle it in the combined distribution later on in this section.

The process can be repeated for Eq.~\eqref{Eq:TransitLengthDensity_AB}. Applying the asymptotic form for the modified Bessel function and writing $\beta$ in terms of $\alpha$ as before gives
\begin{align}
  \tilde{f}_{A|B}(\ell,s) =  C_B \left[ 1 + \sqrt{\frac{s - \ell}{k \ell }} \right] 
  \frac{k \alpha}{2 \sqrt{\pi \alpha \sqrt{ k \ell ( s - \ell ) } }} \exp \left[ -\alpha \left( \sqrt{\ell} - \sqrt{ k (s - \ell) } \right)^2 \right]  .
\end{align}
Transforming $\ell \rightarrow u$ given by Eq.~\eqref{Eq:asymptoticForm_uTransformation_Definition} yields
\begin{align}
  \tilde{f}_{A|B}(u,s) = C_B \left[ \frac{ k ( s - \ell(u) ) }{\ell(u) } \right]^{1/4}  \sqrt{ \frac{\alpha}{\pi} } e^{-\alpha u^2}, \quad -\sqrt{ks} \le u \le \sqrt{s}. \label{Eq:fAB_asymptoticSqrt}
\end{align}
Applying the first-order Taylor series expansion to the fourth root term about $u = 0$ gives the result
\begin{align}
  f^*_{A|B}(u,s) = C_B \left[ 1 - \frac{1}{2} \sqrt{ \frac{1+k}{ks} } u \right] \sqrt{ \frac{\alpha}{\pi} } e^{-\alpha u^2}, \quad -\sqrt{ks} \le u \le \sqrt{s},\label{Eq:fAB_asymptoticLinear}
\end{align}
and a normalization coefficient of
\begin{align}
C_B = 2 \left[ \operatorname{erf}\left(\sqrt{k\alpha s}\right)+\operatorname{erf}\left(\sqrt{\alpha s}\right) - \frac{1}{2} \sqrt{ \frac{1+k}{ks} } \cdot \dfrac{ e^{-k \alpha s}  - e^{-\alpha s}  }{\sqrt{{\pi \alpha}}} \right]^{-1} . \label{Eq:fAB_asymptoticLinear_NormalizationCoefficient}
\end{align}
This is identical to the analogous result for $f^*_{A|A}$ in Eq.~\eqref{Eq:fAA_asymptoticLinear} except for a minus sign on the first-order $u$ term. In contrast to $f^*_{A|A}$, which is skewed toward higher transit lengths, for $f^*_{A|B}$ the particle begins in material type $B$ and this skews the Gaussian negatively toward lower transit lengths in material type $A$. As with its analog, Eq.~\eqref{Eq:fAB_asymptoticLinear} has the potential to become negative. This occurs when $k < 1/3$ and $u > 2 \sqrt{ ks / (k + 1 ) }$. The same considerations discussed previously apply to this case as well.

Equations~\eqref{Eq:fAA_asymptoticLinear} and~\eqref{Eq:fAB_asymptoticLinear} can be combined to form a distribution weighted by the probabilities of starting within a given material type. This becomes
\begin{align}
  f^*_A(u,s) = C \left( 1 + b u \right) \sqrt{ \frac{\alpha}{\pi} } e^{-\alpha u^2}, \quad u_- \le u \le u_+ , \label{Eq:fA_asymptoticLinear}
\end{align}
where
\begin{align}
  b = \frac{ 2 \omega - 1 }{ 2 }  \sqrt{ \frac{1+k}{ks} } \label{Eq:fA_asymptoticLinear_bCoefficient}
\end{align}
and $C$ is a renormalization coefficient. Here the bounds $u_-$ and $u_+$ are chosen to truncate the distribution should it become negative. These are
\begin{subequations}
\begin{align}
  u_- &= \left\{ \begin{array}{l l}
   \mathrm{max} \left\{ -\sqrt{ks}, -\dfrac{1}{b} \right\}, 	& \quad b > 0, \\
   -\sqrt{ks}, 													& \quad b \le 0, \\ \end{array} \right. \label{Eq:fA_asymptoticLinear_uminus} \\
  u_+ &= \left\{ \begin{array}{l l}
    \sqrt{s},												& \quad b \ge 0, \\
    \mathrm{min} \left\{  \sqrt{s}, -\dfrac{1}{b} \right\},	& \quad b < 0 . \\ \end{array} \right. \label{Eq:fA_asymptoticLinear_uplus}
\end{align}
\end{subequations}
The renormalization coefficient is then
\begin{align}
  C = 2 \left[ \mathrm{erf}( \sqrt{\alpha} u_+ ) - \mathrm{erf}( \sqrt{\alpha} u_- ) 
  				- \frac{b}{\sqrt{\pi \alpha}} \left( e^{-\alpha u_+^2} - e^{-\alpha u_-^2} \right) \right]^{-1} .
\end{align}

The use of the first-order Taylor series expansion to linearize the one-fourth root term about $u = 0$ to obtain the asymptotic form introduces an error away from the peak. Stated in reverse, the asymptotic distribution is most accurate near the peak of the distribution where the density is significant. Normally these errors are not a practical concern because one would need to evaluate the distribution far from peak where such contributions would be very small anyway. However, there are potential issues here that may arise. In particular, its use to derive an analytical solution for the transmission of charged particles in thick materials can lead to problems, and this issue is discussed in greater detail in Sec.~\ref{Sec:ChargedParticleStoppingPorousMedia_Transmission}.

\subsection{Atomic-Mix Limit} \label{Sec:TransitLengthDistribution_AtomicMixLimit}

Next, we demonstrate that these expressions satisfy the atomic-mix limit. We begin with the more accurate Eq.~\eqref{Eq:fAA_asymptoticSqrt}, prior to taking the Taylor series expansion. The goal is to show that the transit-length distribution tends toward a Dirac delta function about $u = 0$, which, from Eq.~\eqref{Eq:l_at_u_equal_zero} is the atomic-mix limiting value of the expected distance. One standard definition of the Dirac delta function is a Gaussian limiting toward zero width and infinite height:
\begin{align}
  \delta(z) = \lim_{\sigma \rightarrow 0} \frac{1}{\sqrt{\pi}|\sigma|} e^{-z/\sigma^2} . \label{Eq:diracDeltaDefinition}
\end{align}
Letting $\sigma^2 = 1/\alpha$, we have
\begin{align}
 \tilde{f}_{A|A}(u,s) = C_A \left[ \frac{\ell(u) }{ k ( s - \ell(u) ) } \right]^{1/4} \frac{1}{\sqrt{\pi}|\sigma|} e^{-u^2/\sigma^2}, \quad -\sqrt{ks} \le u \le \sqrt{s}. 
\end{align}
Taking the limit as $\sigma \rightarrow 0$ and applying the definition of the Dirac delta function in Eq.~\eqref{Eq:diracDeltaDefinition} gives
\begin{align}
 f_{A|A}^\infty(u,s) = C_A \left[ \frac{\ell(u) }{ k ( s - \ell(u) ) } \right]^{1/4} \delta(u), \quad -\sqrt{ks} \le u \le \sqrt{s}. 
\end{align}
Using the identity
\begin{align}
  g(z) \delta(z) = g(0) \delta(z) , 
\end{align}
we have
\begin{align}
 f_{A|A}^\infty(u,s) = C_A \left[ \frac{\ell(0) }{ k ( s - \ell(0) ) } \right]^{1/4} \delta(u) = C_A \delta(u), \quad -\sqrt{ks} \le u \le \sqrt{s} ,
\end{align}
and using Eq.~\eqref{Eq:l_at_u_equal_zero},
\begin{align}
  \left[ \frac{\ell(0) }{ k ( s - \ell(0) ) } \right]^{1/4} = 1 . \nonumber
\end{align}
Clearly, the renormalization coefficient $C_A$ must equal one since $u = 0$ is in the domain and
\begin{align}
  \int_{-\infty}^\infty \delta(z) dz = 1. 
\end{align}

Equation~\eqref{Eq:fAA_asymptoticLinear}, the form following the Taylor series expansion, also satisfies the atomic-mix limit. This can be shown using the same process. Additionally, the normalization coefficient $C$ given by Eq.~\eqref{Eq:fAA_asymptoticLinear_NormalizationCoefficient} also goes to one, as expected. Noting that $\mathrm{erf}(z) \rightarrow 1$ as $z \rightarrow \infty$ means the first and second terms both tend toward one, adding to two. The third term tends toward zero, so in the limit we have $C = 2/2 = 1$.

The process for showing that the two cases of $f_{A|B}$, Eqs.~\eqref{Eq:fAB_asymptoticSqrt} and~\eqref{Eq:fAB_asymptoticLinear}, go to the atomic-mix limit is very similar and therefore not repeated.

\subsection{Implementations} \label{Sec:TransitLengthDistribution_Implementation}

Three different implementations of the the transit-length distributions were written in the following languages: C++, Python, and Matlab. Each of these only account for the continuous (or non-Dirac delta function) portion of the density in Eq.~\eqref{Eq:TransitLengthDensity_A}. The Dirac delta function terms at the extrema $\ell = s$ and $\ell = 0$ are excluded from these routines because the density is infinite at that point and in a discontinuous manner. Therefore, when the distribution is applied during integration, the user must exercise special care to explicitly evaluate the Dirac delta function terms and add those to the results from the numerical quadrature. The reason for excluding the Dirac delta function terms is that including infinities would pose issues with interpolation during quadrature. We provide Matlab implementations using these routines applied to porous media (details are discussed in the Secs.~\ref{Sec:LightTransmissionPorousMedia} and~\ref{Sec:ChargedParticleStoppingPorousMedia}) illustrating how this is handled.

\subsubsection{Numerical Challenges}

Evaluating the transit-length distribution in Eq.~\eqref{Eq:TransitLengthDensity_A} poses some numerical challenges regardless of the choice of programming language. The boundaries need special treatment prescribed by Eqs.~\eqref{Eq:TransitLengthDensity_A_limit_0} and~\eqref{Eq:TransitLengthDensity_A_limit_s}. The argument of the Bessel function can become too large such that numerical evaluation of those terms yields a number larger than the largest representable double-precision floating point number. This occurs when the argument $z$ exceeds about 700. Therefore, when $z \ge 700$, the algorithm switches to the asymptotic forms in Eq.~\eqref{Eq:fA_asymptoticLinear}, which are numerically stable for large argument about the peak.

As mentioned, Eq.~\eqref{Eq:TransitLengthDensity_A} has Dirac delta functions at $\ell = s$ and $\ell = 0$ and are not handled explicitly at these extremes. Again, the Dirac delta function terms need to be handled manually. 

\subsubsection{C++ Implementation}

The C++ implementation uses the C++17 standard, taking advantage of the built-in modified Bessel function routines in the C++ Standard Template Library. Two cases are defined in a header and implementation file. The first is an implementation that takes input arguments as doubles of the evaluation point $\ell$, length $s$, mixing parameters $\alpha$ and $\beta$, and volume fraction $\omega$. The second implementation, located in the header file, is templated to take any Standard Template Library container of $\ell$ values, returning a vector of function evaluations. This second implementation calls the first (single point evaluation). A user would insert the header and compile the implementation file within a larger code project.

\subsubsection{Python Implementation}

The Python implementation is a function using the Python3 standard. The dependencies are the math and special functions library in SciPy. The latter of these contain the modified Bessel functions, whereas all other functions are available in the former. The implementation takes any standard Python sequence (tuple, list, etc.) of evaluation points $\ell$ along with individual values of $s$, $\alpha$, $\beta$, and $\omega$ and returns a Python list of evaluations of the transit-length density.

\subsubsection{Matlab Implementation}

The Matlab implementation is a function using standard capabilities that have been available since at least 2006, and therefore should work with any reasonably modern version of the software. The function takes a vector (1-D array) of of evaluation points $\ell$ along with individual values of $s$, $\alpha$, $\beta$, and $\omega$ and returns a vector of evaluations of the transit-length density.

\subsection{Numerical Results} \label{Sec:TransitLengthDistribution_NumericalResults}

We provide numerical evidence for the distribution. First, we verify that the mathematical model and the equations obtained is indeed consistent and the methods of computing the transit-length distribution has been correctly implemented. This is done using Monte Carlo sampling of the telegraph process. The Monte Carlo sampling code is written in C++ and is available on the software repository. Second, we show numerical evidence that the asymptotic form behaves as expected in that it converges to the exact form in the limit of large mixing. This is done using the Matlab version to take advantage of the software's built-in numerical integration routines.

\subsubsection{Monte Carlo Verification} \label{Sec:TransitLengthDistribution_MonteCarloVerification}

The transit-length distribution was derived with a model that postulated a form identical to telegrapher's equation, describing a random process where states switch at exponentially distributed times. A method for verifying this model is to use a Monte Carlo simulation to sample the transit length $\ell$ given a total distance $s$. First a particle begins at $x = 0$ in material $A$ with probability $p_A = \omega$ and material $B$ with probability $p_B = 1 - \omega$. When the particle is in either material $A$ or $B$, the distance to transitioning is sampled using an exponential distribution with rate coefficients given by $\alpha$ and $\beta$. This is
\begin{align}
  d_k = \left\{ \begin{array}{l l} 
  -\ln(\xi_k) / \alpha, & \quad x \in A, \\
  -\ln(\xi_k) / \beta,  & \quad x \in B. \\ \end{array} \right.
\end{align} 
Here $d_k$ is the distance in the $k$th zone to the next transition and $\xi_k$ is the $k$th uniform random number $[0,1)$. If the particle is in material $A$, a score for the transit length of
\begin{align}
    \mathrm{min} \left\{ d_k, s - x \right\} , \quad x \in A , \nonumber
\end{align}
is made. The particle then moves forward by incrementing $x$ by $d_k$. If the new location is greater than $s$, the process terminates. Otherwise, the material type switches from $A$ to $B$ or vice versa and the process repeats until the termination criterion is met.

\begin{figure}[htb!]
\begin{center}
\includegraphics[scale=0.8,trim=0cm 6cm 0cm 6cm]{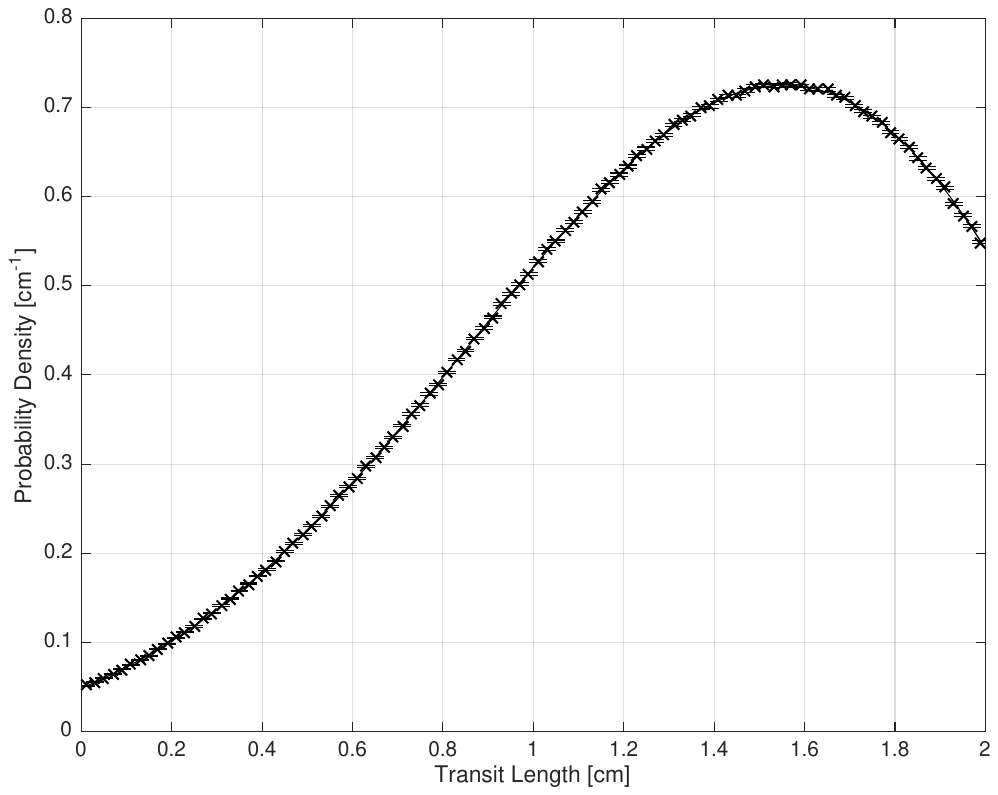}
\caption{Verification of the Transit-Length Density with Monte Carlo Sampling}
\label{Fig:TransitLengthDistribution_MonteCarloVerification}
\end{center}
\end{figure}

The test case uses $s = 2$~cm, $\alpha = 1$~cm$^{-1}$, $\beta = 2$~cm$^{-1}$, and $\omega = 2/3$. The Monte Carlo calculation was run using $10^9$ random samples. The continuous part of the density function (not the Dirac delta terms) using 100 uniform bins is plotted in Fig.~\ref{Fig:TransitLengthDistribution_MonteCarloVerification}. The markers denote the Monte Carlo result and the curve denotes the evaluation of the function. The error bars are the 1-$\sigma$ confidence band. The results agree within these statistical confidence intervals.

The Dirac delta function term for $\ell = 0$, corresponding to the particle starting in and streaming through material type $B$ entirely, gives a probability of $6.10521 \times 10^{-3}$. The Monte Carlo result is $6.10407 \times 10^{-3} \pm  2.463 \times 10^{-6}$. The term for $\ell = s$, for starting in and streaming through material $A$ entirely is $9.02235 \times 10^{-2}$. The Monte Carlo calculation gives $9.02297 \times 10^{-2} \pm 9.060 \times 10^{-6}$. These both agree within 1$\sigma$. This provides numerical evidence that the model and its implementation are correct.

\subsubsection{Error of the Asymptotic Form}

The asymptotic form in Eq.~\eqref{Eq:fA_asymptoticLinear} should become an increasingly accurate representation of the exact distribution in Eq.~\eqref{Eq:TransitLengthDensity_A} as the degree of mixing increases (i.e., larger $\alpha$ and $\beta$). The error between the distribution can be quantified using the L-1 norm,
\begin{align}
  \epsilon_1 = \int_0^s | f_A(\ell,s) - f_A^*(\ell,s) | d\ell.
\end{align}
Higher-order norms, such as the more conventional L-2 or Euclidian norm, diverge. This is because the distribution of the asymptotic form is infinite at the bounds while its integral is finite, but squaring (or taking a higher power) of the difference yields an infinite integral.

\begin{figure}[htb!]
\begin{center}
\includegraphics[scale=0.8,trim=0cm 6cm 0cm 6cm]{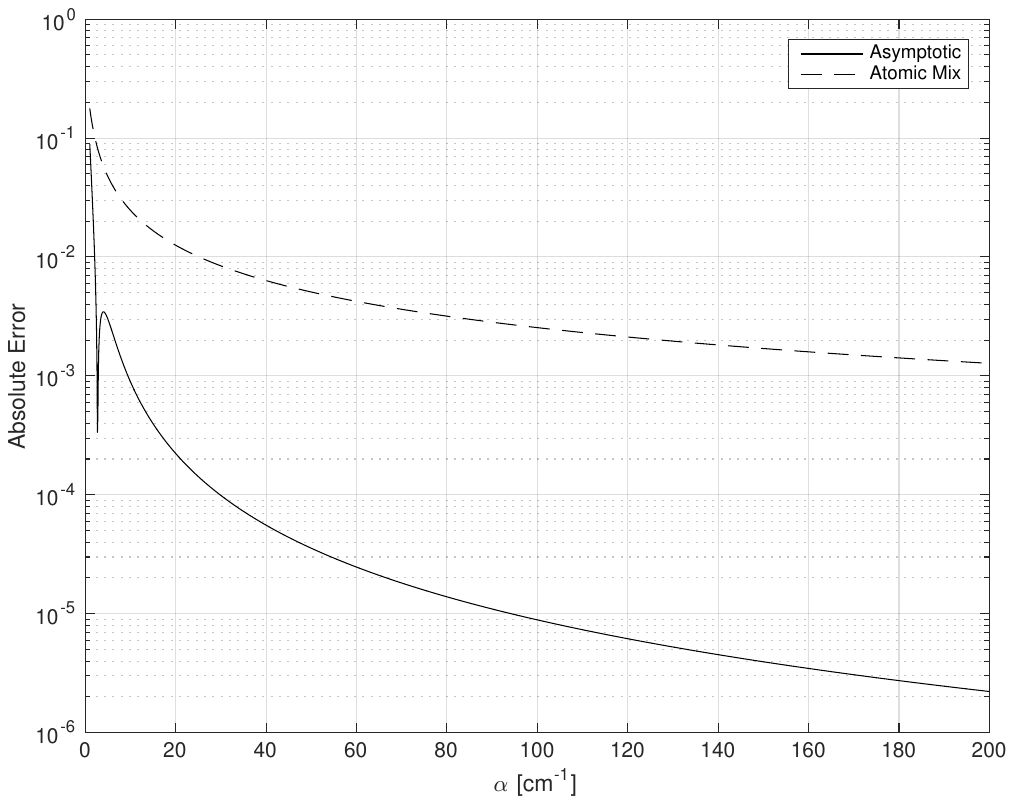}
\caption{L-1 Error of the Exact and Asymptotic Transit-Length Density}
\label{Fig:TransitLengthDistribution_L1Error}
\end{center}
\end{figure}

To demonstrate that the L-1 norm diminishes for larger $\alpha$, we provide numerical evidence a case $s = 2$~cm, $\omega = 1/3$ and variable $\alpha$ with $\beta = 5\alpha$. The L-1 norm is computed shown as a function of $\alpha$ on a log-log plot in Fig.~\ref{Fig:TransitLengthDistribution_L1Error}. This includes only the density term and neglects the Dirac delta function terms, which vanish in the limit of large $\alpha$. The results shows that the asymptotic form approaches (in an integral sense) the exact form in the limit of large $\alpha$ per expectations. Furthermore, its slope on the log-log plot shows that the error (given by the L-1 norm) goes asymptotically as $\alpha^{-1}$.

\clearpage
\section{Transmission of Light Through Porous Media} \label{Sec:LightTransmissionPorousMedia}
In this section, we apply the transit-length distribution to the transport of light through porous media. The results herein, however, are applicable to any neutral particle in a purely absorbing medium. 

The problem contains a sequence of random regions of material and void in slab geometry using Markovian statistics. There are two options for orienting the random material zones. One is to orient the separating planes along the $x$-axis in a layer-cake like geometry where each layer has a random thickness. The other is to orient the zones along the direction of flight, which is more representative of a porous medium and therefore used in this section. The only difference between the two is that the rate parameters $\alpha$ and $\beta$ need to be scaled by $\mu$ in the random layer-cake geometry, whereas they are not scaled in the model where the randomness is always taken along the direction of flight. 


\subsection{Exact Solution}
Suppose the problem is light incident upon a 1-D slab of width $X$ containing a random mixture of material $A$ that is a pure absorber and material $B$ that is void. The locations of material regions $A$ and $B$ are prescribed by a sequence of exponentially distributed random variables with parameters $\alpha$ and $\beta$ respectively. The volume fraction of material $A$ is
\begin{align}
  \omega = \frac{\Lambda_A}{\Lambda_A + \Lambda_B}.
\end{align}
The quantity $1 - \omega$ is the porosity or the volume fraction that is material $B$ or void. Let $\ell$ be the random length of material $A$ between 0 and $x$ described by the transit-length distribution in Eq.~\eqref{Eq:TransitLengthDensity_A} with $s = x/\mu$ where $\mu$ is the cosine of the angle with respect to the $x$-axis.

The transport of light though this problem may be modeled by the 1-D, slab-geometry linear Boltzmann equation without scattering,
\begin{align}
  &\mu \dho{\psi}{x} + \Sigma_t(x) \psi(x,\mu,\ell) = 0, \quad \psi(0,\mu) = g(\mu), \ 0 \le \mu \le 1, \quad \psi(X,\mu,\ell) = 0, \  -1 \le \mu \le 0,
\end{align}
on the domain $0 \le x \le L$, $-1 \le \mu \le 1$. Here $\Sigma_t(x)$ is a random attenuation coefficient given by
\begin{align}
  \Sigma_t(x) = \left\{ \begin{array}{l l}
  \Sigma_t, & \quad x \in A, \\
  0,		& \quad x \in B, \end{array} \right.
\end{align}
where $\Sigma_t$ is a constant. The angular flux $\psi(x,\mu,\ell)$ is also a random variable conditional on the random distance or transit length $\ell(x)$ through material $A$. The boundary condition is assumed to have the same angular distribution $g(\mu)$ regardless of the starting location and that the probability of starting in either region is equal to its volume fraction, which is motivated by physical considerations.

The solution to the transport problem can be obtained as
\begin{align}
  \psi(x,\mu,\ell) = g(\mu) \exp \left[-\frac{1}{\mu} \int_0^x \Sigma_t(x') dx' \right] = g(\mu) e^{-\Sigma_t \ell / \mu}.
\end{align}
We can compute the ensemble average of the angular flux over all random realizations with
\begin{align}
  \psi(x,\mu) = \int_0^{x/\mu}  f_A(\ell,x/\mu) g(\mu) e^{-\Sigma_t \ell / \mu} d\ell. \label{Eq:angularFlux_porousMediaExact}
\end{align}
The ensemble-average leakage or transmission rate is given by the rightward partial current at $x = X$. This is
\begin{align}
  J^+(X) = \int_0^1 \mu \psi(X,\mu) d\mu. \label{Eq:rightPartialCurrent}
\end{align}

While Eq.~\eqref{Eq:angularFlux_porousMediaExact} is an exact solution for the ensemble-averaged angular flux, it unfortunately does not have a form in terms of standard functions and therefore must be determined numerically. It is, however, possible to find an approximate form using the asymptotic forms of the transit-length distribution that are reasonably accurate under the suitable conditions.

\subsection{Approximate Solution}
If the slab is suitably thick such that the particle is expected to encounter a large number of material transitions while traversing the slab, then asymptotic form given by Eq.~\eqref{Eq:fA_asymptoticLinear} is valid. 

The exponential attenuation factor in Eq.~\eqref{Eq:angularFlux_porousMediaExact} poses an obstacle to arrive at an analytical solution; however, this can be addressed in an approximate manner with a Taylor series expansion about $u = 0$, where $u$ is the transformation given by Eq.~\eqref{Eq:asymptoticForm_uTransformation_Definition} where $u = 0$ corresponds to the atomic-mix limiting value of the transit length. This permits an analytic solution because integrals of the form
\begin{align}
  \int u^n e^{-u^2} du \nonumber
\end{align}
can be expressed using standard functions.

Here we take a second-order Taylor series expansion of the exponential attenuation factor about $u = 0$,
\begin{align}
  e^{-\Sigma_t u} \approx e^{- \omega \Sigma_t x / \mu} \left[ 1 - \frac{2 \Sigma_t \sqrt{ k ( 1 + k ) x/\mu } }{ ( 1 + k )^2 } u + \frac{ \Sigma_t ( k^2 + 2 k \Sigma_t (x / \mu) - 1 ) }{ ( 1 + k )^3 } u^2 \right]. \label{Eq:exponentialAttenuation_secondOrderApprox_u}
\end{align}
The validity of this assumption requires that the exponential does not change too rapidly with respect to the significant range of the Gaussian peak given by $e^{-\alpha u^2}$. In other words, the quality of this approximation degrades with large optical thickness, but improves with smaller mean chord lengths such that the width of the Gaussian is narrower and the atomic-mix limit becomes more valid. In principle, more terms in the expansion could be used to improve this approximation and the final result still has an analytic form, but would be significantly more complicated.

Inserting the asymptotic transit length distribution, Eq.~\eqref{Eq:fA_asymptoticLinear}, and the second-order expansion for the exponential attenuation factor, Eq.~\eqref{Eq:exponentialAttenuation_secondOrderApprox_u}, into the exact expression for the ensemble-average angular flux in Eq.~\eqref{Eq:angularFlux_porousMediaExact} and then grouping terms yields an expression of the form:
\begin{align}
  \psi(x,\mu) = C g(\mu) e^{- \omega \Sigma_t x / \mu} \int_{u_-}^{u_+}  \left( 1 + c_1 u + c_2 u^2 + c_3 u^3 \right) \sqrt{ \frac{\alpha}{\pi} } e^{-\alpha u^2} du . \label{Eq:asymptoticAngularFlux_unevaluted}
\end{align}
The coefficients are given by
\begin{subequations}
\begin{align}
  a_1 &= \frac{2 \Sigma_t \sqrt{ k ( 1 + k ) x / \mu } }{ ( 1 + k )^2 }, \\
  a_2 &= \frac{ \Sigma_t ( k^2 + 2 k \Sigma_t (x/\mu) - 1 ) }{ ( 1 + k )^3 } , \\
  c_1 &= b - a_1, \\
  c_2 &= a_2 - a_1 b, \\
  c_3 &= a_2 b.
\end{align}
\end{subequations}
The value of $b$ is from Eq.~\eqref{Eq:fA_asymptoticLinear_bCoefficient} and the limits of integration given in Eqs.~\eqref{Eq:fA_asymptoticLinear_uminus} and~\eqref{Eq:fA_asymptoticLinear_uplus} except that $s = x/\mu$.

The integral in Eq.~\eqref{Eq:asymptoticAngularFlux_unevaluted} can be evaluated in an indefinite sense as
\begin{align}
  \Psi(t) = \frac{ 2 \alpha + c_2 }{ 4 \alpha } \mathrm{erf}{\left( \sqrt{\alpha} t \right)} 
       - \frac{ \alpha ( c_1 + c_2 t ) + c_3 ( 1 + \alpha t^2 ) }{ 2 \sqrt{\pi } \alpha^{3/2} } e^{-\alpha t^2 } .
\end{align}
Therefore, the ensemble-averaged angular flux is
\begin{align}
  \psi(x,\mu) = C g(\mu) e^{- \omega \Sigma_t x / \mu}  \left[  \Psi \left ( u_+ \right) - \Psi \left( u_- \right) \right] . \label{Eq:angularFlux_porousMediaApproximate}
\end{align}
The estimated leakage or transmission rate may then be evaluated by inserting this result into Eq.~\eqref{Eq:rightPartialCurrent}. Unfortunately, there is no known analytical form, so this must be evaluated numerically.

To review, we derived an exact form for the angular flux, which is given by Eq.~\eqref{Eq:angularFlux_porousMediaExact}, that must be evaluated with numerical integration and an approximate analytical solution, which is given by Eq.~\eqref{Eq:angularFlux_porousMediaApproximate}. The approximation relies on a few simplifications. First, the asymptotic form of the modified Bessel function is applied to the transit-length distribution, which is valid in the case of large argument. The resulting form is that of a skewed Gaussian. Second, a first-order Taylor series is applied to linearize a term that scaled as the one-fourth power; this is valid so long as the Gaussian is sufficiently peaked such that the value of the distribution is negligible near the bounds of the distribution, implying that the particle traverses numerous random regions. Finally, a second-order Taylor series expansion is applied to the exponential attenuation. This is valid so long as the exponential does not change too rapidly over the peak of the Gaussian, which means that the material is not too optically thick. The validity of these assumptions is demonstrated numerically in the next section.

\subsection{Numerical Results}
This section presents a series of numerical results showing transmission rates for various scenarios. All results in this section were generated using Matlab scripts, which are available in the software repository. First, we provide results for the transmission probability for two cases using the exact, asymptotic, and atomic-mix models as a function of the degree of mixing. The first case is a normally incident beam source of light, and the second is for an isotropic flux boundary condition. Then, we show how the exact model behaves as parameters are varied and quantify the accuracy of the asymptotic and atomic-mix models. The results show that for media with a high degree of mixing, the asymptotic model is more accurate than the atomic-mix model. 

\subsubsection{Transmission Probabilities}

The case considered is a slab with parameters $X = 1$~cm, $k = 0.2$, $\Sigma_t = 6$~cm$^{-1}$. With these parameters, the slab has an optical thickness of one mean-free-path in the atomic-mix limit. The choice of $k$ means this slab is quite porous with a volume fraction of material of $1/6$. The two cases for boundary conditions are: a beam source,
\begin{align}
  g_1(\mu) = \delta( 1 - \mu ),
\end{align}
and an isotropic flux or cosine source,
\begin{align}
  g_2(\mu) = 2, \quad 0 \le \mu \le 1.
\end{align}
The transmission probabilities are computed as a function of $\alpha$, where higher values of $\alpha$ for fixed $k$ correspond to a higher degree of mixing.

\begin{figure}[htb!]
\begin{center}
\includegraphics[scale=0.8,trim=0cm 6cm 0cm 6cm]{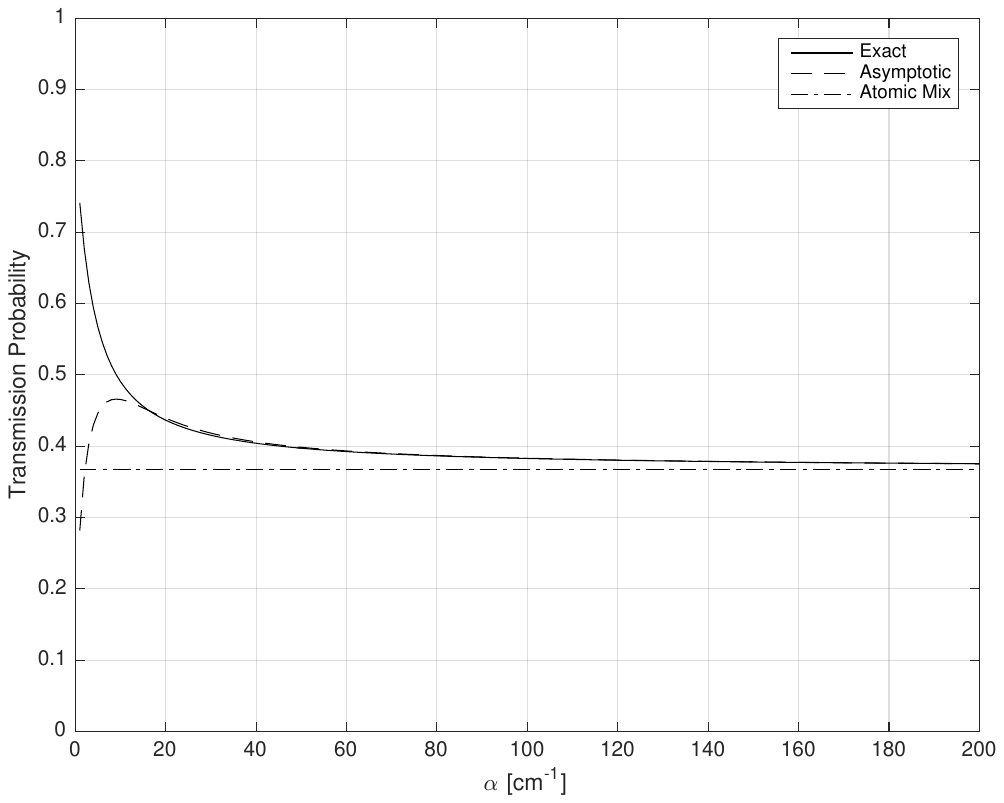}
\caption{Transmission Probability of Porous Slab with a Beam Source}
\label{Fig:PorousTransmission_BeamSource}
\end{center}
\end{figure}

\begin{figure}[htb!]
\begin{center}
\includegraphics[scale=0.8,trim=0cm 6cm 0cm 6cm]{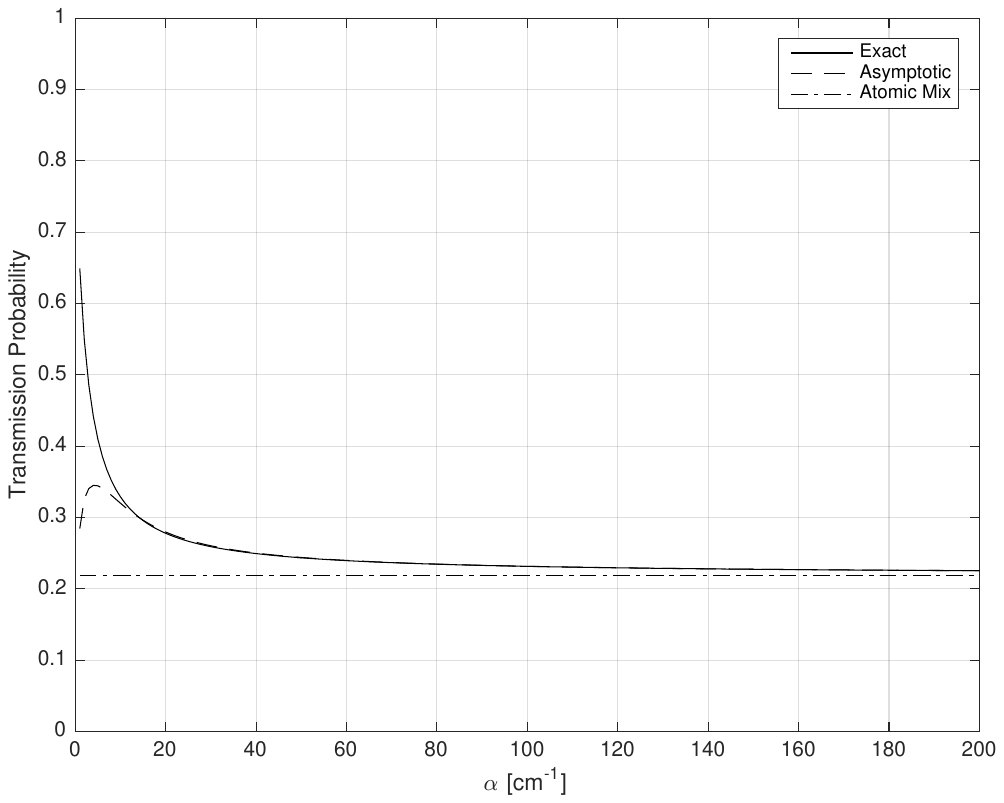}
\caption{Transmission Probability of Porous Slab with an Isotopic Flux Boundary Condition}
\label{Fig:PorousTransmission_IsotropicFlux}
\end{center}
\end{figure}

Figures~\ref{Fig:PorousTransmission_BeamSource} and~\ref{Fig:PorousTransmission_IsotropicFlux} show the transmission probabilities computed using the various models for the two respective cases, beam source and isotropic flux. Note that in cases where the modified Bessel function argument becomes too large in the exact model, the asymptotic transit-length distribution with the exact (not Taylor expanded) exponential attenuation factor is used; this only arises for small values of $\mu$, which are a very minor contribution to the overall transmission probability. For the isotropic-flux case, a grid of 100 even cosine $\mu$ intervals are used to compute the angular flux and the trapezoid rule is used to perform the numerical integration.

The results show that the asymptotic model converges to the exact model when $\alpha$ becomes large. In the beam source case, the error becomes less than 0.1\% when $\alpha$ is about 90~cm$^{-1}$. The atomic-mix model converges very slowly to the exact model, achieving an error of 0.1\% at an $\alpha$ value exceeding 4000~cm$^{-1}$. A more thorough assessment of the error is discussed in the next section.

While not presented here, the trend of the asymptotic model converging to the atomic-mix model holds for a large range of parameters studied. This suggests that the asymptotic model is an accurate and inexpensive method to compute the transmission probability.

\subsubsection{Parameter Study of Model Accuracy}

\begin{figure}[htb!]
\begin{center}
\includegraphics[scale=0.8,trim=0cm 6cm 0cm 6cm]{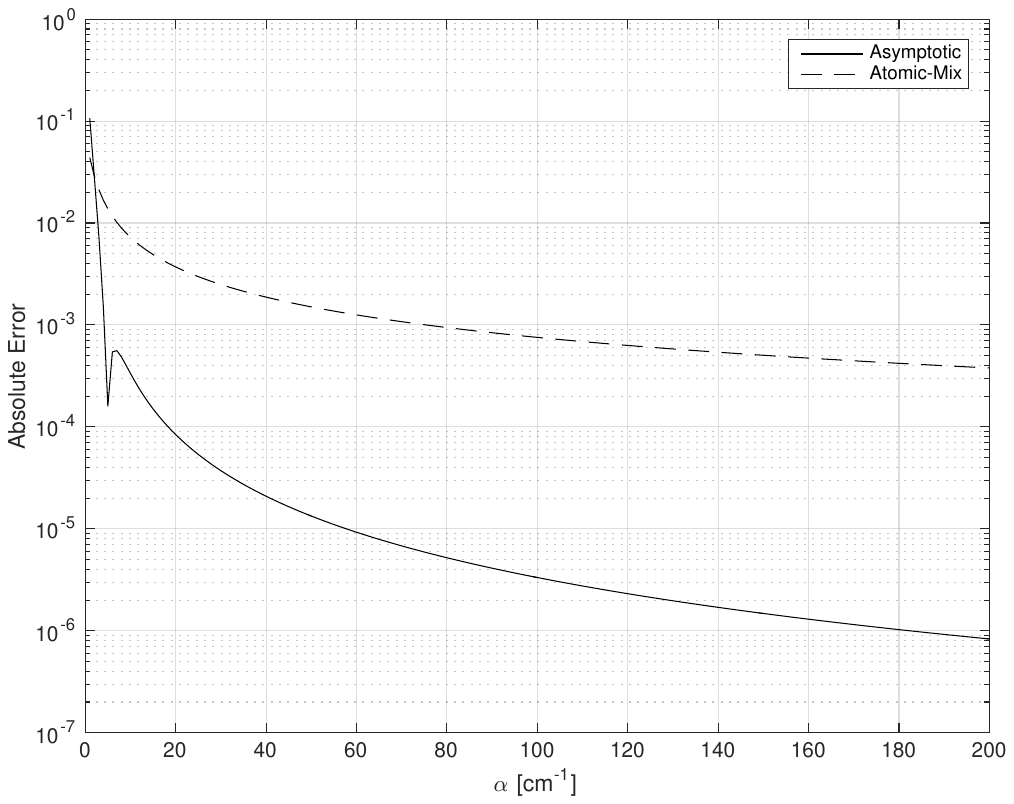}
\caption{Error of Transmission Probability for Various Degrees of Mixing}
\label{Fig:PorousTransmission_ErrorMixing}
\end{center}
\end{figure}

\begin{figure}[htb!]
\begin{center}
\includegraphics[scale=0.8,trim=0cm 6cm 0cm 6cm]{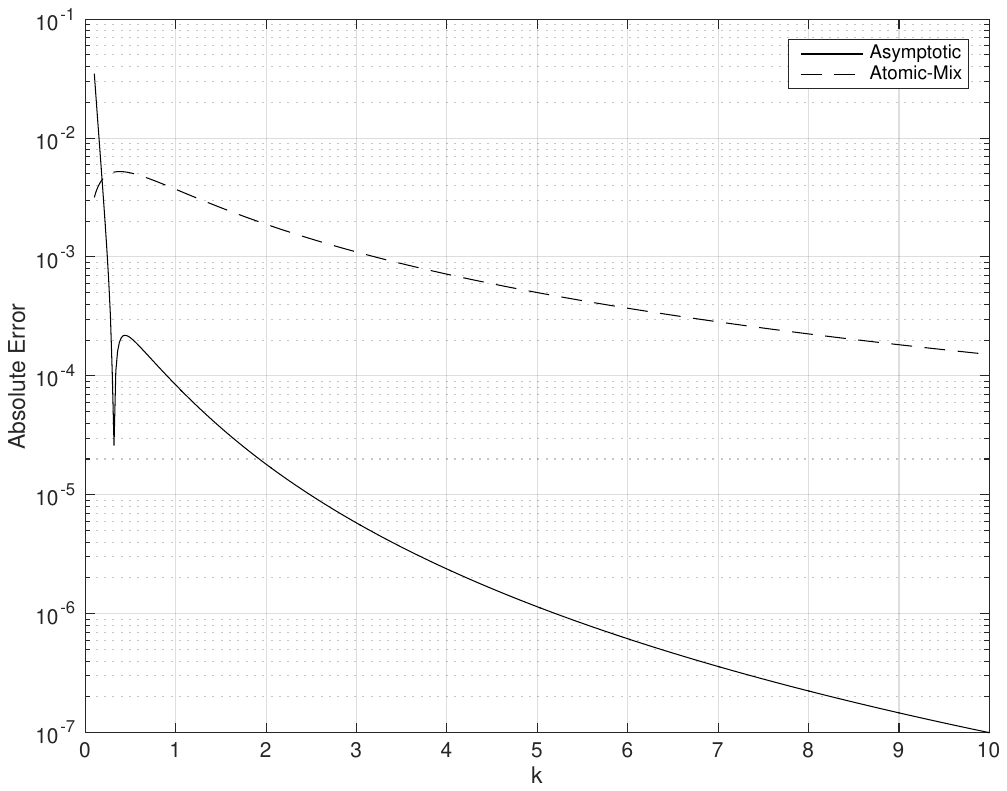}
\caption{Error of Transmission Probability for Various Porosities}
\label{Fig:PorousTransmission_ErrorPorosity}
\end{center}
\end{figure}

\begin{figure}[htb!]
\begin{center}
\includegraphics[scale=0.8,trim=0cm 6cm 0cm 6cm]{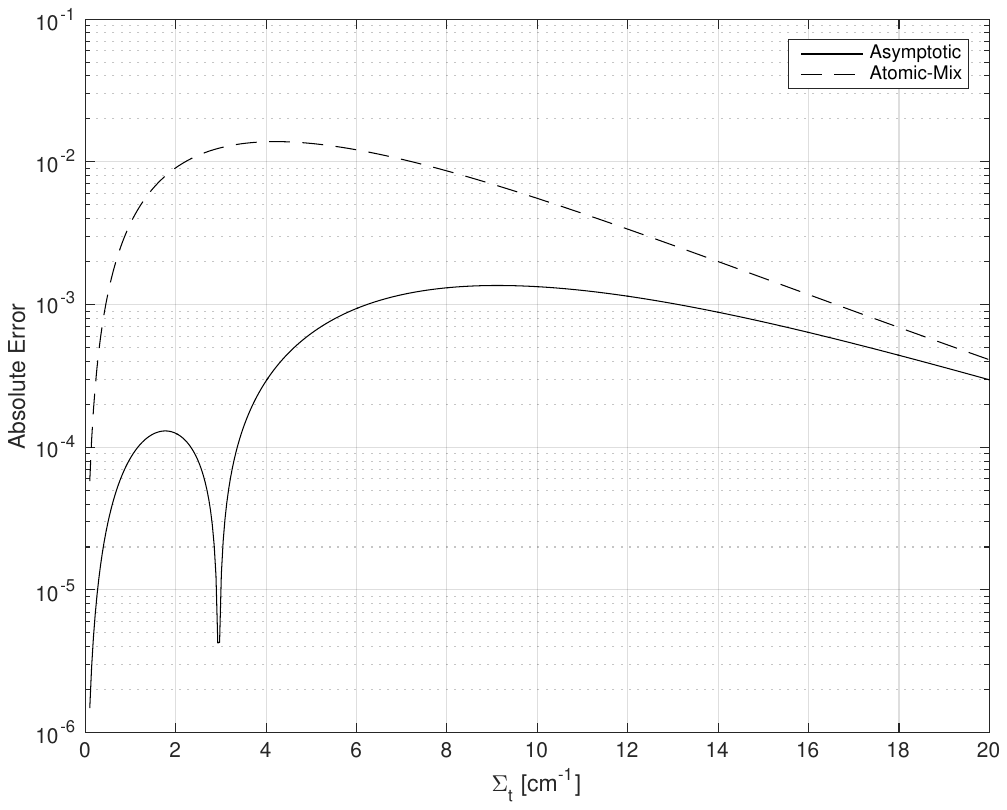}
\caption{Error of Transmission Probability for Various Absorption Cross Sections}
\label{Fig:PorousTransmission_ErrorAbsorption}
\end{center}
\end{figure}

\begin{figure}[htb!]
\begin{center}
\includegraphics[scale=0.8,trim=0cm 6cm 0cm 6cm]{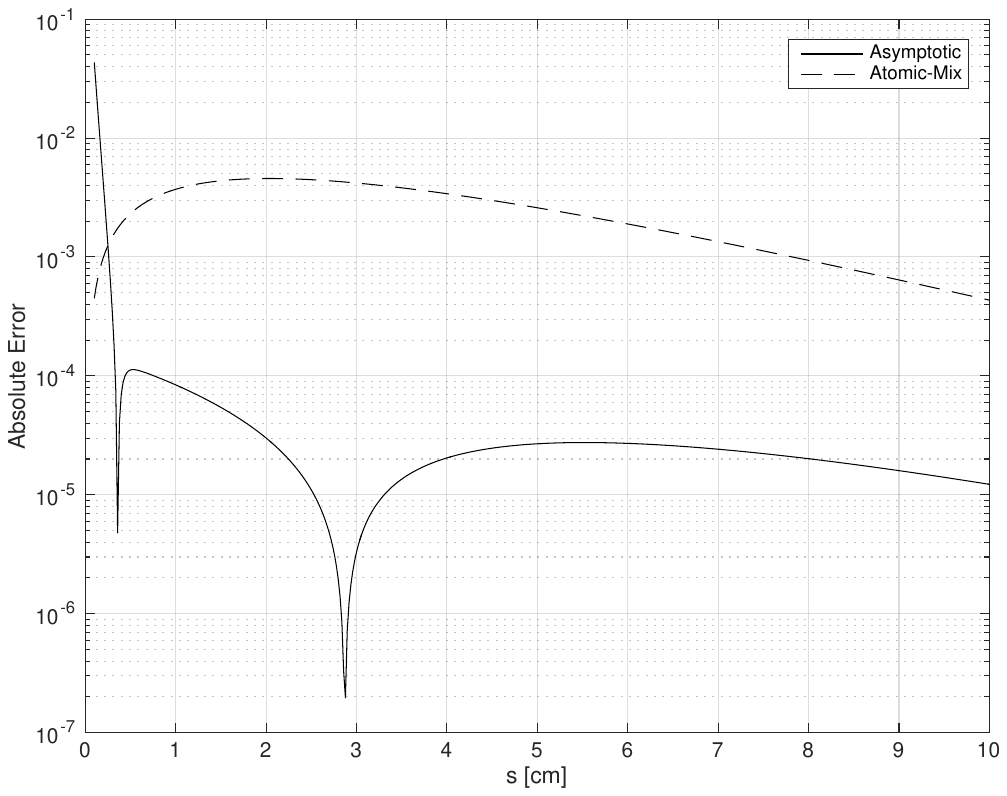}
\caption{Error of Transmission Probability for Various Slab Thicknesses}
\label{Fig:PorousTransmission_ErrorThickness}
\end{center}
\end{figure}

We now study the accuracy of the asymptotic and atomic-mix model as functions of rate coefficient $\alpha$, complement of the porosity $k$, the macroscopic cross section $\Sigma_t$, and the slab thickness $X$. The reference case is $\alpha = 20$~cm$^{-1}$, $k = 1$, $\Sigma_t = 1$~cm$^{-1}$, and $X = 1$~cm. All cases use a beam source. (Additional testing using a different source distribution was performed, but did not yield different conclusions. Therefore, these cases are not presented here.) The absolute error for each of the parameters are presented, in the given order, in Figs.~\ref{Fig:PorousTransmission_ErrorMixing} through~\ref{Fig:PorousTransmission_ErrorThickness} on a semi-log scale. Any jumps occur because the sign of the error switches. In all cases, the asymptotic model outperforms the atomic-mix model in regimes where the asymptotic model is reasonable. This trend is not specific to the case presented here and appears to be a fairly robust result for a large range of parameters.

Figure~\ref{Fig:PorousTransmission_ErrorMixing} gives the absolute error as a function of the degree of mixing $\alpha$. As expected from the results presented in the previous section, the asymptotic model is more accurate than the atomic-mix model for large $\alpha$. For the cases studied, the asymptotic model tends to overestimate the transmission probability in this regime.

Figure~\ref{Fig:PorousTransmission_ErrorPorosity} presents the absolute error as a function of porosity where smaller values of $k$ mean a more porous material. The asymptotic model performs poorly for highly porous materials, which is expected because the argument of the modified Bessel function is not particularly large, violating one of the model's assumptions. The model becomes more accurate as the porosity decreases (larger $k$).

Figure~\ref{Fig:PorousTransmission_ErrorAbsorption} displays the absolute error for varied macroscopic cross section or absorptivity of the slab. The error has a different trend in the previous results. While the asymptotic model is always more accurate the error in both models increases as $\Sigma_t$ becomes bigger but then diminishes for large $\Sigma_t$. For optically thin cases, there is very little absorption and the mixing has little impact on the transmission coefficient. For the asymptotic model, there is a complex interplay of the error because of the accuracy of the second-order Taylor series expansion.

Figure~\ref{Fig:PorousTransmission_ErrorThickness} gives the absolute error as a function of slab thickness. The trend is much the same as the other parameters, where the models become more accurate for a thicker slab.

\clearpage

\clearpage
\section{Charged Particle Stopping in Porous Media} \label{Sec:ChargedParticleStoppingPorousMedia}
This section investigates the stopping of charged particles with a particular focus on electrons. In our analysis, we use a simple model based solely on the stopping power, neglecting important features such as energy straggling, angular deflection, and the production of secondary photons and electrons through hard collisions and bremsstrahlung, respectively. Here we use the same problem as in the previous section: a slab, having thickness $X$, of randomly oriented regions of material and void mixed along the particle direction of flight. Electrons begin within each material type with a probability equal to its volume fraction.

\subsection{Transmission Probability} \label{Sec:ChargedParticleStoppingPorousMedia_Transmission}

The slowing down of electrons is described by the stopping power for a material,
\begin{align}
  S(E) = -\frac{dE}{dx},
\end{align}
which includes energy losses from both collisions and bremsstrahlung radiation. There are various functional forms for the stopping power valid over particular energy ranges, but here we will assume it is tabular data. 

Suppose an electron begins with energy $E_0$ and slows down to energy $E$. The distance the electron travels through the material can be related to this energy loss via the stopping power by
\begin{align}
  \ell = \int_E^{E_0} \frac{dE'}{S(E')} , \quad 0 \le E \le E_0. \label{Eq:relate_distance_energyLoss}
\end{align}
The expected range $R$ of an electron transporting through matter can be found by letting the final energy $E = 0$ (or the some effective lower bound of interest) such that
\begin{align}
  R = \int_0^{E_0} \frac{dE'}{S(E')} .
\end{align}

We consider electrons moving through a slab a distance $s = x/\mu$, where $\mu$ again is the cosine of the angle between the direction of flight and the $x$-axis. The transmission probability is
\begin{align}
  T(x,\mu) = \displaystyle\int_0^R f_A(\ell,x/\mu) \, d\ell .
\end{align}
Inserting in the exact transit-length distribution, Eq.~\eqref{Eq:TransitLengthDensity_A}, we obtain the following result for the transmission probability:
\begin{align}
  T(x,\mu) &= \omega e^{-\alpha x/\mu} \Theta( R - x/\mu ) + ( 1 - \omega ) e^{-\beta x/\mu} \nonumber \\ 
  &+ \left[ \omega \alpha + (1 - \omega) \beta \right] \displaystyle\int_0^R e^{-\alpha \ell - \beta ( x/\mu - \ell )} I_0 \left( 2 \sqrt{\alpha \beta \ell (x/\mu - \ell ) } \right) \, d\ell \nonumber \\
  &+ \sqrt{\alpha \beta} \displaystyle\int_0^R \frac{ ( 2 \omega - 1 ) \ell + ( 1 - \omega ) x/\mu }{ \sqrt{ \ell ( x/\mu - \ell ) }  }  e^{-\alpha \ell - \beta ( s - \ell )} I_1 \left(  2 \sqrt{\alpha \beta \ell (x/\mu - \ell ) } \right) \, d\ell .
\end{align}
The first term accounts for particles that transport through a single region of the material. Transmission occurs only when the chord length is less than the range, $x/\mu < R$. The second term accounts for the case where the electron streams through void without encountering a material. The third and fourth terms involve odd and positive-even number of material transitions as discussed before. These integrals must be done numerically for the exact transit-length distribution. 

Note that when $x/\mu < R$, all particles moving in that direction will transport the entire distance $s$ regardless of whatever random mixing occurs, as the particle has enough energy to transmit through even if it encounters no voids, and the result will always yield one. Therefore, this may be written as an equivalent piecewise function:
\begin{align}
  T(x,\mu) = \left\{ \begin{array}{l l}
    1, & \  x/\mu < R, \\
  ( 1 - \omega ) e^{-\beta x/\mu} + \left[ \omega \alpha + (1 - \omega) \beta \right] & \\
  \times \displaystyle\int_0^R e^{-\alpha \ell - \beta ( x/\mu - \ell )} I_0 \left( 2 \sqrt{\alpha \beta \ell (x/\mu - \ell ) } \right) \, d\ell & \\ 
  + \ \sqrt{\alpha \beta} \displaystyle\int_0^R \frac{ ( 2 \omega - 1 ) \ell + ( 1 - \omega ) x/\mu }{ \sqrt{ \ell ( x/\mu - \ell ) }  }  e^{-\alpha \ell - \beta ( s - \ell )} & \\
  \times I_1 \left(  2 \sqrt{\alpha \beta \ell (x/\mu - \ell ) } \right) \, d\ell , & \ x/\mu \ge R. \end{array} \right. \label{Eq:chargedTransmission_exact}
\end{align}

The transmission probability for a chord can be used to construct the angular flux
\begin{align}
  \psi(x,\mu) = g(\mu) T(x,\mu),
\end{align}
where $g(\mu)$ is the angular flux prescribed at the boundary. The transmission rate of the distribution of electrons is given by the outgoing partial current given by Eq.~\eqref{Eq:rightPartialCurrent}.

There are three regimes of interest for the transmission. Here we consider the simple case of a beam source with $g(\mu) = \delta( 1 - \mu )$. The first, as discussed, is where the thickness of the slab is less than the range such that all electrons transmit through. For the second case, $X/\mu > R$, there are two limiting behaviors with respect to how well mixed the porous material is. If the chord length $s$ is greater than the range $R$, but less than the range divided by the volume fraction of the material $R/\omega$, then the electron transmission probability will go to one in the atomic-mix limit. Likewise, if the chord length $s$ is greater than $R/\omega$, then the transmission probability will go to zero in the atomic-mix limit. The special case where $s = R/\omega$ limits to 1/2.

\subsubsection{Approximate Asymptotic Solution} 

It is also possible to obtain an analytical result with the asymptotic form. Care must be taken in evaluating the integral because of the truncation of the distribution that avoids negative probability density. Inserting Eq.~\eqref{Eq:fA_asymptoticLinear}, this is, for $x/\mu \ge R$,
\begin{align} \label{Eq:chargedTransmission_asympotic}
  T(x,\mu) &= \frac{1}{2} \left[ \mathrm{erf}( \sqrt{\alpha} u_* ) - \mathrm{erf}( \sqrt{\alpha} u_- ) 
  				- \frac{b}{\sqrt{\pi \alpha}} \left( e^{-\alpha u_*^2} - e^{-\alpha u_-^2} \right) \right] \Theta( u_* - u_- ).
\end{align}
Here $b$ is from Eq.~\eqref{Eq:fA_asymptoticLinear_bCoefficient} and $u_-$ is the lower bound including the possibility of truncation from Eq.~\eqref{Eq:fA_asymptoticLinear_uminus} with $s = x/\mu$. Additionally, $u_*$ is the upper bound in that considers the particle range and possible truncation; this is
\begin{align} 
  u_* &=  \left\{ \begin{array}{l l}
  \sqrt{R} - \sqrt{k(x/\mu - R)},													& \quad b \ge 0, 	\\
  \mathrm{min} \left\{  \sqrt{R} - \sqrt{k(x/\mu - R)}, -\dfrac{1}{b} \right\} .	& \quad b < 0. 		\\ \end{array} \right.
\end{align} 
The Heaviside step function is present to handle the case where the particle range is smaller than the location of truncation on the negative side. The transmission for the case where $x/\mu < R$ is one.

As mentioned in Sec.~\ref{Sec:TransitLengthDistribution_AsymptoticForms}, the accuracy of this approximate solution is limited by the use of the first-order Taylor series expansion to linearize a term about the atomic-mix limiting transit distance. This implies the underlying distributions are accurate about the peak, but suffer from errors far away from it. In this particular application, the integration is from zero to the range $R$. This integration domain includes the peak only when $x/\mu < R/\omega$, i.e., the chord length is less than the range and particles transmit in the atomic-mix limit. Therefore, when $x/\mu > R/\omega$, the integration domain only includes portions of the distribution for which the use of the Taylor series approximation is questionable. When this form is applied in the $x/\mu < R/\omega$ case, the relative error decreases with increased mixing. Conversely, when it is applied to the $x/\mu > R/\omega$ case, the relative error does not diminish.

In summary, this asympotic solution should only be used for the case where the particles transmit in the atomic-mix limit. In principle, one could address this by taking the first-order Taylor series about a different point (perhaps half the range), but this idea is not explored here.

\subsection{Energy Spectrum} \label{Sec:ChargedParticleStoppingPorousMedia_EnergySpectrum}

The energy spectrum of the transmitted electrons is important to calculate quantities such as biological dose. By using the stopping power, the expected particle energy following transmission can be computed using Eq.~\eqref{Eq:relate_distance_energyLoss} for a given distance $\ell$ traveled through a material. Because of the porosity in the problem being considered, the distance traveled is now a random variable given by the transit-length distribution, which leads to a spread in the energy loss. Note that this is distinct from fluctuations in the energy loss because of the inherent randomness of the electron interactions with the background material called energy straggling. We do not consider energy straggling in this paper, but instead focus exclusively on the randomness caused by the background material.

Any electron that travels a distance less than its range, $\ell < R$, is not stopped. The energy-loss distribution for these electrons with an initial energy $E$ at depth $x$ along direction cosine $\mu$ can be obtained using a variable transformation with
\begin{align}
  \frac{d\ell}{dE} = \frac{1}{S(E)}
\end{align}
and is
\begin{align}
  \Phi(E,x/\mu) = \frac{f_A(E,x/\mu)}{S(E)}  , \quad 0 \le E < E_0.
\end{align}
Either the exact or asymptotic forms of the transit-length distribution may be used as appropriate. Note that the integral over all $E$ is the transmission probability $T$ and not unity if $x/\mu > R$, as some of the electrons would need to travel farther than their range in matter and are stopped or lost prior to reaching the position $x$.

The angular flux of electrons can be found by multiplying the distribution by the boundary flux $g(\mu)$. This is
\begin{align}
  \psi(x,\mu,E) = g(\mu) \Phi(E,x/\mu) .
\end{align}
The leakage or transmission rate of particles can be found by the outgoing partial current. This is computed by integrating out all outgoing directions and energies,
\begin{align}
  J^+(X) = \int_0^\infty \int_0^1 \mu \psi(X,\mu,E) d\mu dE .
\end{align}
These integrals must be done numerically.

\subsection{Numerical Results} \label{Sec:ChargedParticleStoppingPorousMedia_NumericalResults}

Numerical results are generated for 100~MeV electrons incident on a porous slab consisting of iron. The stopping power data used in these calculations was generated by MCNP6.2. Matlab scripts were then used to generate results. These scripts and the iron stopping power data file are available in the software repository. The porosity of the slab for all cases is 75\%, such that $k = 0.25$. The estimated range $R$ of these electrons is approximately 3.37~cm. Three sets of results are presented: transmission probabilities, energy spectra, and angular fluxes or currents.

\subsubsection{Transmission Probability}

The first case involves computation of the transmission probabilities for a beam of electrons normally incident $(\mu = 1)$ upon slabs of various thicknesses as a function of $\alpha$, the expected rate that particles transition out of a material zone. Since the porosity is fixed, varying $\alpha$ changes the degree of mixing. Smaller values mean the material is more heterogeneous whereas larger ones increase the degree of homogenization.

\begin{figure}[htb!]
\begin{center}
\includegraphics[scale=0.8,trim=0cm 6cm 0cm 6cm]{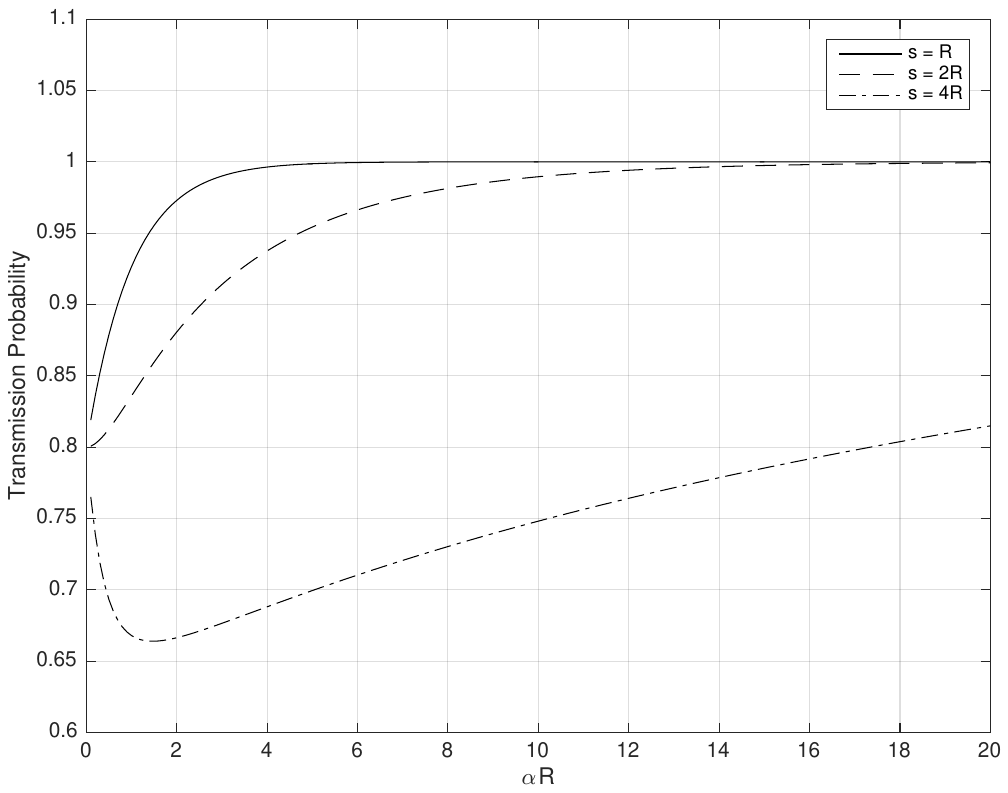}
\caption{Electron Transmission Probabilities for Thin Porous Slabs with Varied Mixing}
\label{Fig:ElectronTransmission_Thin}
\end{center}
\end{figure}

First, we consider cases where the slab thickness is less than the range in the atomic-mix limit, which is $5R$. Figure~\ref{Fig:ElectronTransmission_Thin} shows the transmission probabilities $T$ for the thin slabs with thicknesses $R, 2R$, and $4R$ as a function of $\alpha R$. As expected, the transmission probabilities approach one as the mixing ($\alpha R$) increases. The case with thickness $4R$ initially decreases before increasing toward one. The reason for this is because for low values of $\alpha R$, the transmission is dominated by the case where the electrons stream through a small number of large voids, which are on-average thicker than their material-containing counterparts. These become less probable as $\alpha R$ increases and the electron can be fully stopped by encountering enough thick layers of material. As the mixing given by $\alpha R$ continues to get large encountering enough material to fully stop the electron becomes increasingly unlikely.

\begin{figure}[htb!]
\begin{center}
\includegraphics[scale=0.8,trim=0cm 6cm 0cm 6cm]{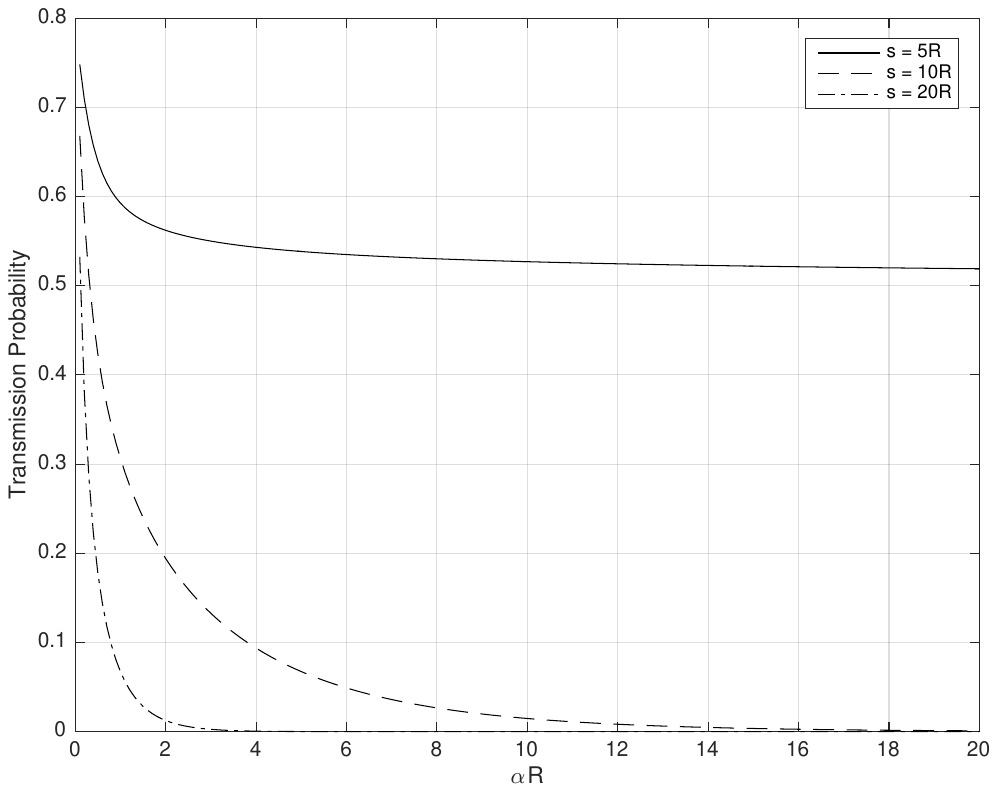}
\caption{Electron Transmission Probabilities for Thick Porous Slabs with Varied Mixing}
\label{Fig:ElectronTransmission_Thick}
\end{center}
\end{figure}

Next, we analyze cases where the slab thickness is equal to or greater than the range in the atomic-mix limit, which we refer to as thick slabs. These are $5R, 10R$, and $20R$ and displayed in Fig.~\ref{Fig:ElectronTransmission_Thick}. For the case where the thickness is $5R$, the value tends toward one half because the transit-length distribution becomes similar to a Gaussian centered around the range with half of the probability on each side. The thicker cases converge to zero as $\alpha R$ increases, as they can both fully stop electrons in the atomic-mix limit.

\begin{figure}[htb!]
\begin{center}
\includegraphics[scale=0.8,trim=0cm 6cm 0cm 6cm]{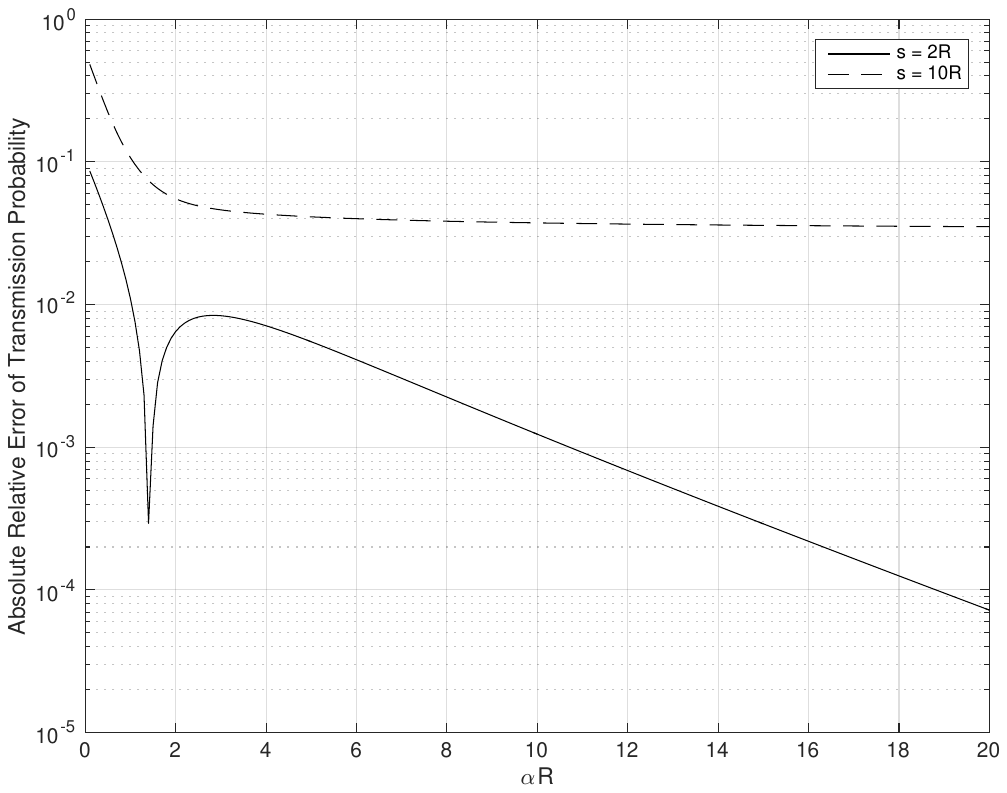}
\caption{Absolute Value of the Error of Asymptotic Versus Exact Models for Electron Transmission}
\label{Fig:ElectronTransmission_AsymptoticError}
\end{center}
\end{figure}

Finally, we test the accuracy of the asymptotic form of the transmission coefficient from Eq.~\eqref{Eq:chargedTransmission_asympotic}. As mentioned earlier, the asymptotic form is expected to perform well when the particles can transmit through the slab in the atomic-mix limit as the mixing increases. Figure~\ref{Fig:ElectronTransmission_AsymptoticError} shows the absolute value of the relative error between the exact and asymptotic solutions for cases for slabs with thicknesses of $2R$ and $10R$ (again $5R$ is the range in the atomic-mix limit). The results show that for a thickness of $2R$, the relative error diminishes as the mixing increases. However, because limitations of the asymptotic form for thick slabs that was previously discussed, the relative error does not decrease for the $10R$-thickness slab and actually begins to grow. (The magnitude of the error does, however, decrease for large $\alpha R$. There is also a sudden dip for small $\alpha R$, which is the result of a sign change in the error.)

\clearpage

\clearpage

\clearpage

\subsubsection{Energy Spectrum}

The second set of results is for a beam of electrons normally incident on the slab with a thickness of $10R$. Nominally, all electrons would be stopped in the atomic-mix limit, but the presence of porosity permits some of those electrons to transmit through the slab. 

\begin{figure}[htb!]
\begin{center}
\includegraphics[scale=0.8,trim=0cm 6cm 0cm 6cm]{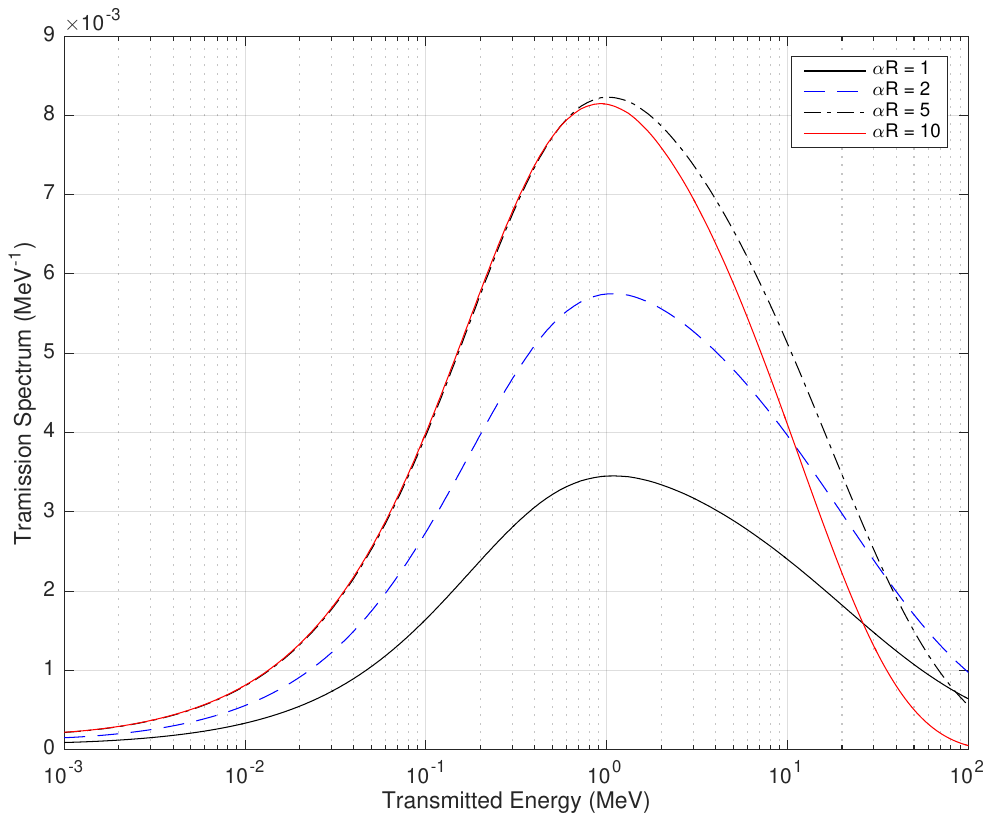}
\caption{Energy Spectra for Transmitted Electrons for Various Mixing Levels}
\label{Fig:ElectronTransmission_Spectra}
\end{center}
\end{figure}

The energy spectra of the transmitted electrons for slabs with various $\alpha$ is presented in Figure~\ref{Fig:ElectronTransmission_Spectra}. Only the continuous part of the spectrum is displayed. Contributions for electrons that transmit through void and experience no slowing down are not shown, since they are a Delta function spike; these probabilities are, respectively, $0.38, 0.18, 0.02$, and $4.2 \times 10^{-4}$. Note that the area under the curve grows initially with increasing $\alpha$, as fewer electrons are able to transmit through without encountering a material and experience some amount of slowing down, but do not get completely stopped. As $\alpha$ continues to increase, the spectrum shifts toward lower energies.

\subsubsection{Angular Flux and Current for an Isotropic Boundary Condition}

The third set of results give the energy-integrated angular flux and outgoing partial current for the slabs of thickness $2R$ with varied $\alpha$, but for the boundary condition of an isotropic flux with $g(\mu) = 2$. 

\begin{figure}[htb!]
\begin{center}
\includegraphics[scale=0.8,trim=0cm 6cm 0cm 6cm]{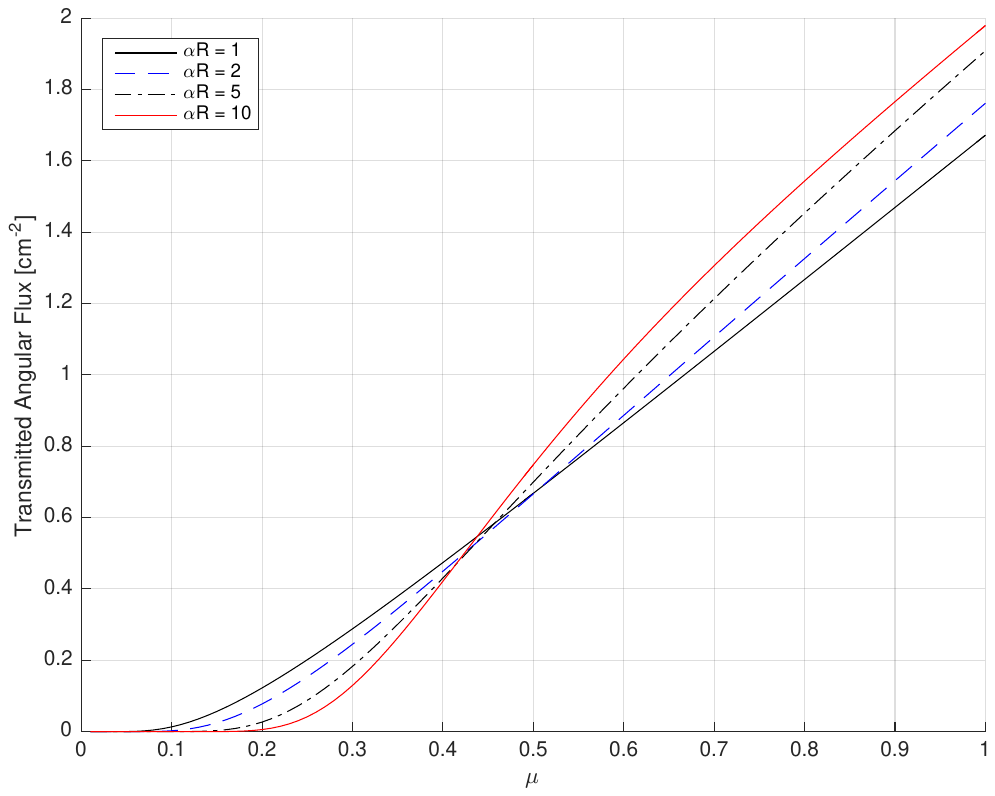}
\caption{Transmitted Angular Flux of Electrons with Isotropic Boundary Condition}
\label{Fig:ElectronTransmission_AngularFlux}
\end{center}
\end{figure}

The energy-integrated angular fluxes $\psi(X,\mu)$ is shown for the various slabs in Fig.~\ref{Fig:ElectronTransmission_AngularFlux} as a function of $\mu$. As $\alpha$ (the amount of mixing) increases, the angular flux becomes more forward peaked. Particles traveling in directions that are increasingly tangent to the slab (lower $\mu$) must transmit through a larger amount of material and are more prone to being stopped, and a higher $\alpha$ means that there is a higher probability that the electron will encounter an amount of material that is at least the range.

\begin{figure}[htb!]
\begin{center}
\includegraphics[scale=0.8,trim=0cm 6cm 0cm 6cm]{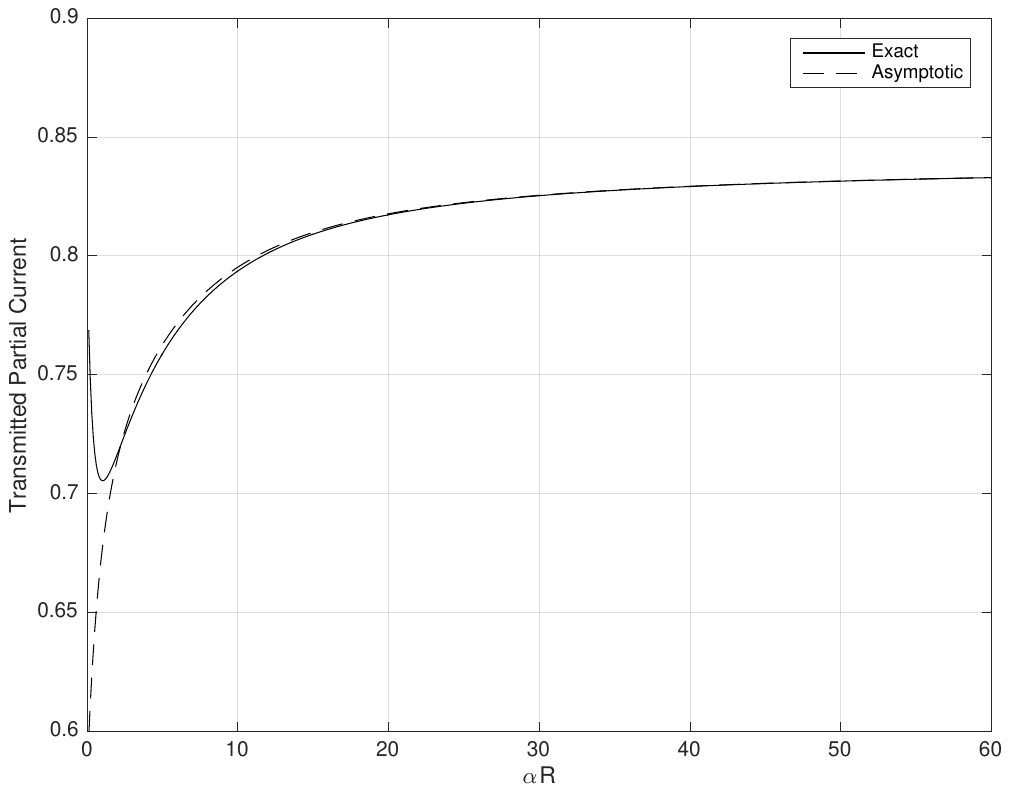}
\caption{Transmitted Partial Current of Electrons with Isotropic Boundary Condition}
\label{Fig:ElectronTransmission_PartialCurrent}
\end{center}
\end{figure}

The outgoing partial current (transmission) is computed as a function of $\alpha R$ using the exact and asymptotic models. These are displayed in Fig.~\ref{Fig:ElectronTransmission_PartialCurrent}. As the amount of mixing increases, the number of electrons transmitting through the slab approaches an asymptotic atomic-mix limiting value, which is consistent with what is expected.

\clearpage

\clearpage

\clearpage

\clearpage
\section{Summary \& Future Work} \label{Sec:Conclusions}
In this paper, we derived the transit-length distribution in binary Markovian media and an asymptotic form in the limit of a high degree of mixing that satisfies the atomic-mix limit. This distribution has been applied to two applications involving porous media: the transmission of light in purely absorbing media and the stopping of charged particles. For the former, we developed an analytical asymptotic form that is more accurate than the atomic-mix approximation.

Future applications of the transit-length distribution may include the development of reaction rate path-length estimators in Monte Carlo simulations of binary Markovian media. It should be possible to construct an estimator that computes the expected or average reaction rate for each simulated particle flight as opposed to a single random realization. Such an estimator would have a lower variance, but would be more costly to compute. Whether such an estimator is more efficient is a subject of further study. While more speculative, it may be possible to use the transit-length distribution to develop sampling schemes and estimators that are either more efficient or capable of estimating higher moments of responses (e.g., the variance because of the stochastic nature of the background material).

Another potential application of the transit-length distribution is to use it to provide homogenized material properties that contain some information about the mixing statistics of particle trajectories. Such a homogenization would contain information about the relative frequency of how much distance a particle trajectory spends in each material type. Indeed, we have already applied the transit-length distribution in this manner to the transport of electrons in a condensed history model. This permits a more accurate sampling of the energy loss, angular deflection, and the emission of secondaries compared to using the atomic-mix approximation.

\section*{Acknowledgments}

This work was partially funded by the Consortium for Monitoring, Technology, and Verification under DOE/NNSA award number DE-NA0003920. The authors would like to acknowledge Dr. Aaron Olson for insightful feedback that improved the quality of the article.


\end{document}